\documentclass[useAMS,usenatbib]{mn2e}
\usepackage{amsmath}
\usepackage{amssymb}
\usepackage{graphicx}
\usepackage{epsfig}
\makeatletter{}
\def\Mpch{\mbox{$h^{-1}$Mpc}}
\def\kpch{\mbox{$h^{-1}$kpc}}

\def\M200{\mbox{$M_{\rm 200 }$}}
\def\Msunh{\mbox{$h^{-1}M_\odot$ }}

\def\R200{\mbox{$R_{\rm 200 }$}}

\def\V200{\mbox{$V_{\rm 200 }$}}

\newcommand{\Ng}{\mbox{$N_{\rm g}$}}

\newcommand{\lsim}{\mbox{${\,\hbox{\hbox{$ < $}\kern -0.8em \lower 1.0ex\hbox{$\sim$}}\,}$}}
\newcommand{\gsim}{\mbox{${\,\hbox{\hbox{$ > $}\kern -0.8em \lower 1.0ex\hbox{$\sim$}}\,}$}}

\def\beqn{\vspace{2mm}
\begin{eqnarray}} 
\def\eeqn{\vspaceg{2mm} 
\end{eqnarray}}

\newcommand{\be}{\begin{equation}}
\newcommand{\ee}{\end{equation}}
\newcommand{\ba}{\begin{eqnarray}}
\newcommand{\ea}{\end{eqnarray}}
\newcommand{\brr}{\begin{array}}
 
\newcommand{\err}{\end{array}}
\newcommand{\bc}{\begin{center}}
\newcommand{\ec}{\end{center}}

\title[PDF of the matter density field]{Density distribution of the cosmological matter field}

\makeatletter{}\author[A.~Klypin et al.]
  {Anatoly~Klypin$^{1}$, Francisco~Prada, Juan Betancort-Rijo$^3$, and Franco D. Albareti$^4$\footnote{'la Caixa'-Severo Ochoa Scholar}
   \vspace{0.2cm}\\ 
  $^1$Astronomy Department, New Mexico State University, Las Cruces, NM, USA\\
}

\author[Klypin et al.]{Anatoly~Klypin$^1$\thanks{E-mail: aklypin@nmsu.edu}, Francisco~Prada$^{2}$,
Juan Betancort-Rijo$^{3,4}$, and Franco D. Albareti$^5$\thanks{``la Caixa''-Severo Ochoa Scholar} \\
 \vspace{-0.2cm}\\
$^{1}$ Astronomy Department, New Mexico State University, Las Cruces, NM, USA\\
$^2$ Instituto de Astrof\'{\i}sica de Andaluc\'{\i}a (CSIC), Glorieta de 
     la Astronom\'{\i}a, E-18080 Granada, Spain \\
$^3$ Instituto de Astrof\'{\i}sica de Canarias, C/Via Lactea s/n, E-38205,
La Laguna, Tenerife, Spain\\
$^4$ Universidad de La Laguna, Dpto. Astrof\'{\i}sica, C/Astrof\'{\i}sico Francisco Sanchez s/n, E-38206 La Laguna, Tenerife, Spain\\
$^5$ Instituto de Fisica Teorica UAM/CSIC, Universidad Autonoma de Madrid, Cantoblanco, E-28049 Madrid, Spain
\\
}

\begin{document}
\maketitle
\label{firstpage}
\begin{abstract}
  The one-point probability distribution function (PDF) of the matter
  density field in the universe is a fundamental property that plays
  an essential role in cosmology for estimates such as gravitational
  weak lensing, non-linear clustering, massive production of mock
  galaxy catalogs, and testing predictions of cosmological
  models. Here we make a comprehensive analysis of the dark matter
  PDF using a suite of $\sim 7000$ $N$-body simulations that covers a
  wide range of numerical and cosmological parameters. We find that
  the PDF has a simple shape: it declines with density as a power-law
  $P\propto\rho^{-2}$, which is exponentially suppressed on both small
  and large densities. The proposed double-exponential approximation
  provides an accurate fit to all our $N$-body results for small
  filtering scales $R< 5\Mpch$ with $rms$ density fluctuations
  $\sigma>1$.  In combination with the spherical infall model that
  works well for small fluctuations $\sigma<1$, the PDF is now
  approximated with just few percent errors over the range of twelve
  orders of magnitude  -- a remarkable example of
  precision cosmology. We  find that at $\sim 5-10\%$ level the
  PDF explicitly depends on redshift (at fixed $\sigma$) and on
  cosmological density parameter $\Omega_m$. We  test different
  existing analytical approximations and find that the often used log-normal
  approximation  is always 3-5 times less
  accurate than either the double-exponential approximation or the
  spherical infall model.
\end{abstract}


\makeatletter{}\section{Introduction}
\label{sec:intro}

The one-point probability distribution function (PDF) of the matter
density field in the universe, and its related statistics the distribution
of galaxy counts, have a long and somewhat patchy history in cosmology
and extragalactic astronomy. It was Edwin Hubble almost a century ago
who found that the counts of about $44,000$ extra-galactic nebulae
distributed over a large area of the sky have a probability
distribution that is not Gaussian but can be approximated by a
log-normal distribution \citep{Hubble1934}. The statistics of galaxy counts in the Lick
survey, in projected cells of size $10'\times 10'$, was studied 
by \citet{SoneiraPeebles1978} who also discovered that the
distribution of the counts is much broader than the Poisson PDF. 

The $rms$ of galaxy counts $\sigma$ in cells of size $R$ is an
integral over the power spectrum of the galaxy distribution
\citep[e.g.,][Sec.36]{Peebles}. As such, in former times, a
count-in-cells analysis of the IRAS redshift galaxy survey was
performed by \citet[][]{Efstathiou1990} who used the counts as a
measure of the two-point clustering statistics on different
scales. Once methods to estimate the correlation function and the
power spectrum were developed and new large-scale galaxy surveys were
available, the count-in-cells as clustering statistics started to play
a secondary role. Higher moments of cell counts depend on correlation
functions of order larger than two. This means that the whole PDF has
information not only on the two-point clustering but also on higher
order statistics, which by itself is very valuable information.

At present, a precise description and modeling of the underlying
matter density distribution - and biasing prescription that connects the
dark matter field with the galaxy distribution - are 
fundamental to extract cosmological information from current and
upcoming large-scale redshift and lensing galaxy surveys
\citep[e.g.,][]{Taruya2002,Takahashi2011,Carron2015,Clerkin2017,Manera2013,Kitaura2016}. For
this reason in the last years there has been a rejuvenated interest in
the cosmic density distribution both from cosmological $N$-body
simulations and galaxy surveys.

\citet{Wild2005} estimated the PDF of galaxies in the 2dF
redshift survey using about 200,000 galaxies. Because of a relatively
small volume, their analysis was done only for large cells of size 10-30
Mpc. They found that the log-normal distribution fits the data
reasonably well, but the noise in the data did not allow them to make
accurate measurements of the PDF. The situation was improved by
\citet{Hurtado2017} using the count-in-cells statistics for galaxies
in the SDSS main sample. They used $\sim 100,000$ galaxies and estimated
the PDF for spheres of radius $R=(8-24)\Mpch$. They found that the
log-normal distribution was very inaccurate (a factor of $\sim 2$ errors) 
for spheres of $R=8\Mpch$. A modification of the log-normal
distribution (called log-normal+bias) somewhat improved the fits, but
it still had $\sim 50\%$ errors at low number of galaxy counts. The negative
binomial distribution was a much better fit for all filtering
scales. At hight redshift, \citet{Bel2016} studied the
count-in-cells distribution of $\sim 30,000$
galaxies in the VIPERS redshift survey with the typical average number
of galaxies per cell of 0.5-5 and spherical cells of radius
$R=(4-8)\Mpch$. They found that the skewed log-normal distribution (a
modification of the log-normal distribution with 4 more free
parameters) was not accurate enough to fit the results of
observations. Instead, they found that the negative binomial 
distribution was much more accurate. Yet, \citet{Clerkin2017} using
the DES science verification data confirmed that the log-normal model
is a good  fit to both the galaxy density contrast and weak lensing
convergence PDFs on scales of $(3-10)$\,Mpc at median redshift $z = 0.3$. In
spite of the fact that at present these seems to be the best
observational results, the errors and noise in the data are still
substantial. 

On the theoretical side the situation is also complicated. There are two
types of approaches: models that start with some dynamical  description of the
non-linear evolution of the density field and proceed to make
predictions of the matter PDF
\citep[e.g.,][]{Juan1991,Bernardeau1994,Kofman1994,Juan2002,
  Ohta2003,LamSheth2008}; and then there are phenomenological approximations that assume a
specific analytical form of the PDF \citep{ColesJones,NegBinom2000,Lee2017,Shin2017},
and find best-fitted parameters of this distribution function to simulation data.

Theoretical models based on the nonlinear dynamics typically use either some
variant of the spherical infall model
\citep[e.g.,][]{Ohta2003,LamSheth2008,Neyrinck2016} or the Zeldovich
approximation \citep[e.g.,][]{Kofman1994,Juan2002}. These models have
made substantial progress and now can make very accurate predictions
for relatively large smoothing scales $R\gtrsim 5\Mpch$ and small
$\sigma\lesssim 1$ \citep{LamSheth2008} giving errors less than
$\sim 10\%$. Not surprisingly, as expected, the models are not very useful and start
to fail at larger $rms$ fluctuations $\sigma\gtrsim 1$
\citep{LamSheth2008,Neyrinck2016}.

One of the disadvantages of the dynamical models is their
complexity. They typically require some manipulation of the linear
power spectrum, analytical approximations for different terms, and can
be quite cumbersome to deal with. This is not a serious impediment to
their use, but it is a nuisance. Simple analytical functions can serve
as an alternative to more complicated dynamical models.

The log-normal distribution is an example of this approach. It was
heavily advocated by \citet{ColesJones} and is often used for
relatively large smoothing scales. There is little justification why
the density distribution function should be log-normal. \citet{ColesJones} argue that under the
  assumption that the divergence of the peculiar velocity field in
  Eulerian coordinates grows as the velocities themselves (as given by
  linear theory) the density field can be expressed as the exponential
  of a Gaussian field. But while their assumption is acceptable for
  the Lagrangian divergence, for the Eulerian one there is an
  additional growth roughly proportional to the cubic root of the
  normalized density.  This leads to a density field that is equal to
  a Gaussian field to the third power, whose PDF is quite different
  from a log-normal distribution. Here is the
main argument of \citet{ColesJones}: ``The lognormal is one of the
simplest ways of defining a fully self-consistent random field which
always has $\rho>0$ and, most importantly, is one of the few
non-Gaussian random fields for which interesting properties are
calculable analytically.''  This says that the PDF should be
log-normal because it can be handled analytically -- hardly a serious
argument.  Another argument is of the same caliber: the
log-normal distribution is well-known and frequently used in other
fields of science \citep{Ohta2003}.  Similar arguments were used for
other phenomenological models \citep{NegBinom2000,Lee2017}.

The only real justification for the existing phenomenological
approximations (including the log-normal) is that they make a fit to
$N$-body results. This is the reason why cosmological $N$-body
simulations are important for the field. In this paper we use a very
large suite of cosmological simulations to produce accurate estimates
of the dark matter distribution functions. Our simulations cover a
wide range of numerical and cosmological parameters. We use the
estimates to test different dynamical models and approximations for the PDF
and to study its dependance on redshift and cosmological parameters.

One of our goals in this paper is to make a comprehensive study of the
different effects that can be associated with $N$-body results
regarding the matter density distribution, such as mass and force
resolution, size of the box, shot noise and cosmic variance. In this
regard we find that systematic errors in the PDF can be important. For
example, noise related to the  discreteness of the density probed by particles is
the leading factor of seriously wrong estimates of the PDF in
underdense regions.

Generation of mock galaxy catalogs provides  a  motivation for our
study of the density distribution function. One needs to produce
thousands of realizations of the dark matter density and velocity
fields. This can be done by carefully tuning parameters of simulations
and limiting their resolution to a fraction of a megaparsec
\citep[see e.g.,][]{COLA,Chuang2015,GLAM}. A biasing prescription then connects
the dark matter with galaxies. This path requires knowledge of the
distribution of dark matter mass on very small scales $100\kpch-1\Mpch$. This is
a challenge because the resolution of these simulations is not
sufficient to resolve individual halos and subhalos making it
difficult to apply existing tools such as Halo Abundance Matching and
Halo Occupation Distribution. A path to solve the problem is to map
dark matter to galaxies using a biasing scheme
\citep[e.g.,][]{Kitaura2016} which requires 
the understanding details of the density
distribution function  and finding limitations to its  estimates.

Unfortunately, only very few studies in the literature provide PDF
results for small smoothing scales $\lesssim 1\,\Mpch$
\citep{Platen2009,Pandey2013,Lee2017}. So, we will make an effort to
study this regime too.  \citet{Platen2009} finds that in
the regime of small smoothing and large density the PDF has a
power-law shape with a slope of $\approx -1.9$, which is similar to
what we find in this work.

This paper is organized as follows. In Sec.~\ref{sec:def} we define
quantities related with the PDF and provide details of some
analytical approximations in Sec.~\ref{sec:LogN}. The spherical infall  and the double-exponential
models are introduced in Sec.~\ref{sec:sph}. Numerical
simulations used in this paper are discussed in
Sec.~\ref{sec:sim}. Sec.~\ref{sec:main} also presents main features of
the PDF. Accuracy of different approximations are discussed in
Sec.~\ref{sec:sphLog}. Summary of results is
given in Sec.~\ref{sec:concl}. Finally, numerical effects are discussed in
Appendix~\ref{sec:cosmic}, and  
Appendix B presents tables of parameters for the double-exponential
approximation.

\makeatletter{}\section{Definitions}
\label{sec:def}
In order to estimate the density distribution function $P(\rho)$ from
$N$-body simulations we split the
computational volume $L^3$, where $L$ is the box size, with a $3D$
mesh of size $N^3_{\rm cell}$ 
 and  use the Cloud-In-Cell (CIC) density assignment scheme to
estimate the density $\rho$ at each grid point of the mesh. The cell size of the
grid $\Delta x =L/N_{\rm cell}$ defines the smoothing length. The
density is normalized to the average matter density $\rho_{av} \equiv
\Omega_m\rho_{\rm cr}$, i.e.

\begin{equation}
\rho \equiv 1+\delta_{\rm NL} = \rho_{\rm DM}/\Omega_m\rho_{\rm cr},
\label{eq:overdensity}
\end{equation}
where $\delta_{\rm NL}$ is the matter density contrast or overdensity. The index
$NL$ highlights the fact that $\rho$ is a non-linear quantity and it
can be distinguished from the density contrast $\delta$ as estimated
by the linear theory. Throughout the paper we use the quantity $\rho$
as ``density'' in spite of the fact that it is really a normalized density -- a
dimensionless quantity as shown in eq.\,(\ref{eq:overdensity}). This is done
for convenience to avoid repeating $1+\delta_{\rm NL}$ in most plots and equations.

The values of density $\rho$ are binned using
logarithmically spaced bins with width
$\Delta\log_{10}(\rho) =0.025-0.050$. The density distribution
function - PDF of the cosmic density field - is then defined as a
normalized number of cells with density in the range $[\rho,\rho+\Delta\rho]$:

\begin{equation}
P(\rho) = \frac{\Delta N_{\rm cell}}{N^3_{cell}\Delta\rho}.
\label{eq:PDF}
\end{equation}
The PDF can have a surprisingly large range of
values. For example, density can
reach values larger than $10^5$ for hundreds of cells when we use a
large mesh of $\sim 3000^3$ cells in high-resolution simulations.  That gives
$P(\rho) \sim 10^{-12}$. At the same time the number of cells at low
densities can be millions for a small density bin leading to a large PDF
value $P(\rho)\gtrsim 1$.  In order to avoid a large dynamical range of quantities, we
typically plot $\rho^2P(\rho)$.

By design, the density distribution function is normalized to have
the total volume and total mass equal to unity:
\begin{equation}
\int^\infty_0 P(\rho)d\rho =1, \quad \int^\infty_0 \rho P(\rho)d\rho =1.
\label{eq:normalization}
\end{equation}
The second moment of $P(\rho)$ gives the $rms$ density fluctuation of
the density field $\sigma$, and as such it is related to the
non-linear power spectrum $P_{NL}(k)$ of density perturbations:
\begin{equation}
  \sigma^2 = \int^\infty_0 (\rho-1)^2 P(\rho)d\rho 
                    = \frac{1}{2\pi^2}\int^{k_{Ny}}_0P_{NL}(k)W^2(k\Delta x)k^2dk,
\label{eq:sigma}
\end{equation}
where $W^2(k\Delta x)$ is the power spectrum of the CIC filter with
the width $\Delta x$, and the integral is truncated at the Nyquist frequency of the
mesh $k_{Ny}=\pi/\Delta x$.

\makeatletter{}\begin{figure}
\centering
\includegraphics[width=0.49\textwidth]
{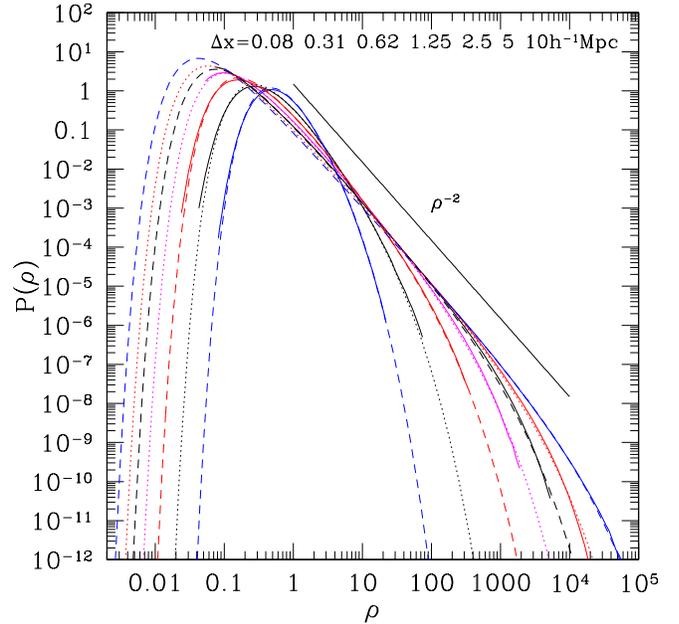}
\caption{Density distribution function at $z=0$ for different
  filtering scales indicated in the plots. Full curves show the results from
  our simulations. Double-exponential models are presented by
  dashed and dotted curves.  The full line shows the power-law behaviour with the slope -2. As
  the filtering scale decreases in value the PDF becomes wider and
  approaches the power-law.}
\label{fig:pdf}
\end{figure}

Figure~\ref{fig:pdf} shows our first results on the structure of the
PDF. More detailed discussion is given later in
Sec.~\ref{sec:main}. The plot demonstrates the main trend of the shape
of the PDF. In the linear regime of the growth of fluctuations, when
$\sigma\ll 1$, the PDF is a Gaussian distribution that quickly acquires
a skewed shape as $\sigma$ increases. With the further increase of
$\sigma$ the PDF becomes progressively wider and becomes a power-law
with the slope close to -2 that is smoothly suppressed on small and large densities.

  \makeatletter{}\section{Methodology}
\label{sec:method}

Different analytical approximations and theoretical models, as mentioned in
Sec.~\ref{sec:intro}, are used to fit and
make predictions for the PDF estimates obtained from numerical
simulations. Here we describe both approaches.

\makeatletter{}\subsection{PDF analytical approximations: log-normal,
  Negative Binomial, and Generalized Extreme Value}
\label{sec:LogN}

We introduce the log-normal, negative
binomial and the generalized extreme value distributions which have
been traditionally adopted as analytical approximations for the PDF in many works.

The {\bf log-normal} (LN) distribution function $P_{LN}$ is defined as:
\begin{equation}
\rho P_{LN}(\rho) = \frac{1}{\sqrt{2\pi\sigma^2_{LN}}}
         \exp\left(-\frac{[\ln(\rho)+\sigma^2_{LN}/2]^2}{2\sigma^2_{LN}} \right),
\label{eq:LN}
\end{equation}
where
\begin{equation}
\sigma^2_{LN} =\ln[1+\sigma^2].
\end{equation}
is the only free parameter. $\sigma^2_{LN}$ can be obtained from
results of simulations, and thus should be considered as fixed.

Because the log-normal distribution does not provide accurate fits to
numerical simulations, a number of modifications have been proposed
\citep{Hamilton1985,Colombi1994,Shin2017}. None of those modifications
extend the approximation to large densities, hence we do not discuss them in this paper.

The {\bf negative binomial} (NBN) distribution
\citep{JuanNB,NegBinom2000,Bel2016} is defined as a discrete distribution. It
is the probability $P_N(V)$ to find $N$ particles in a cell of volume
$V$ with the average number of particles $\bar N$. It can be
re-written as a distribution function of density contrast
$\rho =N/\bar N$:

\begin{equation}
P_{NBN}(\rho) = \frac{1}{\bar N}\cdot \frac{\Gamma(N+1/g)}{\Gamma(1/g)\Gamma(N+1)} 
              \cdot\frac{(g\bar N)^N}{(1+g\bar N)^{N+1/g}},
\label{eq:nbn}
\end{equation}
where  $g$ is a parameter that is defined by $rms$ fluctuations of
counts: $g=(\sigma_N^2-\bar N)/\bar N^2$. The average number of
particles per cell $\bar N$ in general is not an integer number and is
defined by the average density and cell volume: $\bar N = \bar\rho V$.
So, the two parameters $g$ and $\bar N$ that define the NBN distribution are not
free and can be fixed from simulations. However, for PDFs with not too
small $rms$ fluctuations
($\sigma\gtrsim 1$) the fits produced by this distribution are not
very accurate.  As the result, we decided to treat both $\bar N$ and
$g$ as free parameters.

While the negative bimodal distribution formally has two parameters,
there is little change in $P_{NBN}$ when $\bar N>10$. For cases where
NBN makes some reasonable fits, our simulations have typical values of
$\bar N$ of many hundreds. So, in practice the NBN PDF depends
only on one parameter $g$, which defines the width of the distribution
function: the larger is $g$, the wider is $P(\rho)$.

At large average number of objects in a cell $\bar N\gg 1$ and for large
$N>\bar N$ the NBN approximation predicts that the density
distribution changes with density $\rho = N/\bar N$ as follows
\begin{equation}
P_{NBN}(\rho) \approx  \frac{1}{\bar N^{1/g}\Gamma(1/g)} 
              \rho^{1-1/g}\exp\left(-\frac{\rho}{g}\right).
\end{equation} 
This expression is very different compared to the behavior of the PDF
observed in $N$-body simulations for large $\rho$ and $\sigma$: in
that regime
$P(\rho)\propto \rho^{-2}\exp\left(-C\rho^{0.5}\right)$. Thus,
the NBN approximation predicts too steep a decline with density and
lacks the power-law regime of the $N$-body PDF.

A {\bf generalized extreme value} (GEV) distribution as an approximation for
the density distribution function was used by \citet{Lee2017}. It can be written as:
\begin{equation}
\rho P_{GEV}(\rho) = \frac{1}{\ln(10)\beta}\frac{\exp(-z^{-1/k})}{z^{1+1/k}}, \,
      z\equiv 1+\frac{k}{\beta}\lg_{10}\left(\frac{\rho}{\rho_0}\right),
\label{eq:GEV}
\end{equation} 
where $k, \rho_0,$ and $\beta$ are free parameters.

\makeatletter{}\subsection{Spherical Infall and double-exponential PDF models}
\label{sec:sph}

Approximations discussed so far were not based on any dynamical
models.  They simply make a guess regarding the functional form of
$P(\rho)$ and then proceed to finding parameters that produce the best
fit. The guess is not based on any insights from the dynamics of
clustering either. {\bf Spherical infall models} are different because
they are theoretical predictions for the density distribution function
that are based on simplified approximations of the non-linear
evolution of the density field. Here we closely follow the theoretical
framework developed by
\citet{Juan2002} and \citet{LamSheth2008}. We assume that the linear
density field was smoothed with a top-hat filter with radius $R_f$ and
corresponding mass $M$. The variance of the smoothed field is equal to
\begin{equation}
\sigma_L^2(M) = \frac{1}{2\pi^2}\int k^2dk P(k)W^2(kR).
\label{eq:sigmaL}
\end{equation}
In the spherical infall model the mapping from linear density contrast
$\delta_L$ to the nonlinear overdensity $\rho$ is
approximated by the following relation \citep{Juan1991,Bernardeau1994}:
\begin{equation}
\rho = \left(1-\frac{\delta_L}{\delta_c}\right)^{-\delta_c}, \quad
\rho\equiv \frac{M}{\bar M},\quad  \bar M=\rho_b\Delta x^3,
\end{equation}
where $\delta_c$ is the linear theory prediction for the critical
overdensity of collapse. Here we will use $\delta_c=5/3$ as suggested
by \citet{Juan2002}. For the
initial Gaussian fluctuations this model gives the density
distribution function:
\begin{equation}
\rho^2P(\rho) =\frac{1}{\sqrt{2\pi}}\exp\left(-\frac{\delta_L^2}{2\sigma_L^2}\right)
      \frac{d(\delta_L/\sigma_L)}{d\ln\rho}.
\label{eq:sph1}
\end{equation}
This is the same as eq.(6) in \citet{LamSheth2008}. 
We also use a modification of the spherical infall model, which is
based on the excursion set model \citep{Sheth1998,LamSheth2008}:
\begin{equation}
\rho^2P(\rho) =\frac{1}{\sqrt{2\pi\sigma_L^2}}\exp\left(-\frac{\delta_L^2}{2\sigma_L^2}\right)
      \left[1-\delta_L\left(\frac{1}{\delta_c} -\frac{\gamma}{3}\right) \right],
\label{eq:sph2}
\end{equation}
where $\gamma(\rho)$ is:
\begin{equation}
\gamma = -3\frac{d\ln\sigma_L^2}{d\ln M}.
\label{eq:gam}
\end{equation}

Note that in these equations $\sigma_L$ is a function of filtering mass $M$, which
in turn depends on density $\rho = M/\bar M$. So, $\sigma_L = \sigma_L(\rho)$.

In order to apply the models, we need to adjust the top-hat filtering
scale $R_f$ used in the spherical infall model so that it matches the
Cloud-In-Cell filtering scale used in our simulations. This is done by
matching the power spectra of both filters at wavenumbers $k<0.7k_{Ny}$ by
applying $R_f=\Delta x/\sqrt{1.3}$, where $k_{Ny}=\pi/\Delta x$ is the
Nyquist frequency of the density grid used for density assignment, and
$\Delta x$ is the grid size. Specifically, for every bin with
density $\rho$ we find the mass $M=\rho\bar M$ and then the top-hat
filtering radius is found as
$R_f= \sqrt{2}\left[3M/4\pi\rho_{cr}\Omega_m\right]^{1/3}$.
When applying the relations given in
eqs.(\ref{eq:sigmaL}--\ref{eq:gam}), we integrate numerically
eq.(\ref{eq:sigma}) with the top-hat filter $W(kR_f)$ and use finite
differences to estimate the derivatives in eq.(\ref{eq:sph1}) and eq.(\ref{eq:gam}).

\makeatletter{}\begin{table*}
\begin{center}
\caption{Numerical and cosmological parameters of different simulations.
  The columns give the simulation identifier, 
  the size of the simulated box in $h^{-1}\,{\rm Mpc}$,
  the number of particles, 
  the mass per simulation particle $m_p$ in units $h^{-1}\,M_\odot$, the mesh size $\Ng^3$,
  the  gravitational softening length $\epsilon$ in units of $h^{-1}\,{\rm Mpc}$, the number of time-steps $N_s$, 
the amplitude of perturbations $\sigma_8$, the matter density $\Omega_m$,
  the number of realisations $N_r$}
\begin{tabular}{ l | r | c | r |  c|  c | c | c | c | r |r }
\hline  
Simulations & Box\phantom{1} & particles  & $m_p$\phantom{mmm}   & $\Ng^3$  & $\epsilon$ & $N_{\rm s}$ & $\sigma_8$ & $\Omega_m$ & $N_r$ 
\tabularnewline
  \hline 
A0.5         & 500$^3$    & 1200$^3$ & $6.16\times 10^9$   & 2400$^3$ & 0.208 & 181 & 0.822 & 0.307 & 680 & 
\tabularnewline
A1.5         & 1500$^3$   & 1200$^3$ & $1.66\times 10^{11}$ & 2400$^3$ & 0.625 & 136 & 0.822 & 0.307  & 4513 & 
\tabularnewline
A2.5         & 2500$^3$   & 1000$^3$ & $1.33\times 10^{12}$ & 2000$^3$ & 1.250 & 136 & 0.822 & 0.307  & 1960 & 
\tabularnewline
B0.5      & 500$^3$   & 1600$^3$ & $2.66\times 10^{9}$  & 3200$^3$ & 0.156 & 271 & 0.828 & 0.307  & 5 & 
\tabularnewline
D0.5        & 500$^3$   & 1600$^3$ & $2.33\times 10^{9}$  & 3200$^3$ & 0.156 & 271 & 0.828 & 0.270  & 5 & 
\tabularnewline
MDPL1              & 1000$^3$   & 3840$^3$ & $1.5 \times 10^{9}$  & --      & 0.010 & --  & 0.828 & 0.307   & 1 & 
\tabularnewline
BolshoiP        & 250$^3$ & 2048$^3$ & $1.5 \times 10^{8}$  & --      & 0.001 &  -- & 0.828 & 0.307   & 1 & 
\tabularnewline
\hline
\end{tabular}
\label{table:simtable}
\vspace{-5mm}
\end{center}
\end{table*}

As was found previously
\citep[e.g.,][]{Juan2002,LamSheth2008,LamShethEllips2008,
  Neyrinck2016}, the spherical infall model provides good
approximations for the PDF in those cases with large filtering scales where
the $rms$ fluctuation in the simulation box $\sigma(M)$ is relatively
small $\sigma\lesssim 1$.  This is consistent with our results, which
will be presented later. At larger $\sigma$ (large densities) the model provides results
that are much less accurate. Even so, the spherical infall model predicts trends
that track our $N$-body results. This is somewhat unexpected
because at large overdensities $\rho \gtrsim 10^2$ 
the model is clearly outside of the limits of its dynamical
applicability. After all, a simplistic treatment of non-linear
evolution used by the model cannot be valid for densities that are
appropriate for collapsed and virialized halos.

It is interesting and instructive to find what makes the spherical
infall model much better than expected.  The last factor in
eq.~(\ref{eq:sph1}) is just a correction to the leading terms that are
the power-law $P\propto \rho^{-2}$ (see also Figure~\ref{fig:pdf}) and
the exponential term on the right-hand-side of eq.~(\ref{eq:sph1}). The
exponential term originates from the assumption that the density
distribution function of primordial fluctuations is Gaussian. The
$\rho^{-2}$ term comes from integrating the Gaussian distribution
function over mass and then by writing it in differential form. This
is basically the same logic as in the Press-Schechter derivation of
the mass function of dark matter halos.

\begin{figure}
\centering
\includegraphics[width=0.49\textwidth]
{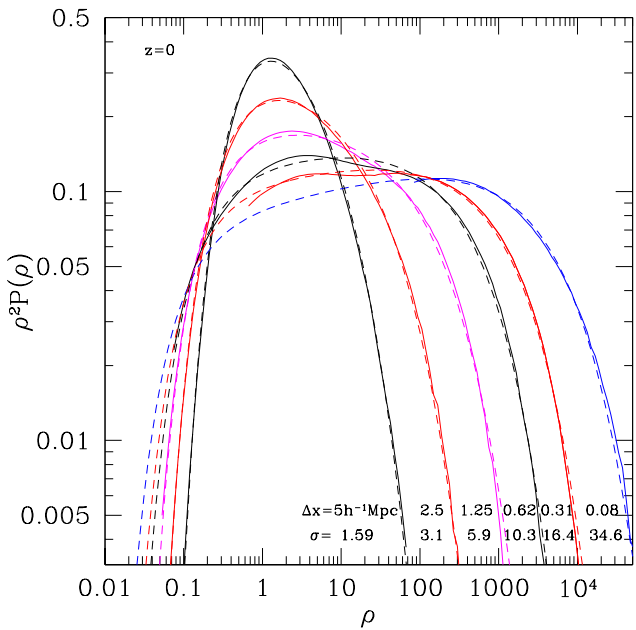}
\caption{Density distribution function scaled with $\rho^2$ at $z=0$ for different
  filtering scales indicated in the plot. Full curves show the results from
  our simulations. Double-exponential models are presented by
  dashed and dotted curves.  As
  the filtering scale decreases in value the PDF becomes wider and
  approaches the power-law. }
\label{fig:pdf2}
\end{figure}

More in detail, we see that the exponential term provides the truncation
of the $P\propto \rho^{-2}$
behavior both on the low-density $\rho<1$ and on the high-density
$\rho\gg 1$ regimes. At low densities (small masses, large $\sigma_L$)
the truncation is mostly due to large negative values of
$\delta_L$. At large $\rho$ (large masses and small $\sigma_L$) the
decline is related to the combination of decreasing $\sigma_L$ and
increasing $\delta_L$. However, the increase of $\delta_L$ is limited:
it can not exceed $\delta_{L,{\rm max}}=\delta_c=5/3$, and the decline
in $\sigma_L$ is not strong enough by itself to produce a substantial
suppression of the PDF. So, the $\rho^{2}P(\rho)$ shape becomes nearly flat
at very large densities $\rho > 100$ and small smoothing scales
$\Delta x \lesssim 1\Mpch$. This is clearly seen in Figure~\ref{fig:pdf2}.

At very small smoothing scales $\Delta x \ll 1\Mpch$ and large
densities there is another regime, that sheds light on how the density
distribution function should behave at very large densities. When
density is larger than $\sim 100$ we are likely dealing with interiors
of collapsed dark matter halos.  In this regime the distribution
function $P(\rho)$ is a sum over the distribution functions of individual
halos. Assuming that the density profile of a halo can be approximated by
the Navarro-Frank-White (NFW) profile, and for a very small filtering
scale, we can can derive the PDF provided by a single
halo. If $\rho(r)$ is the halo density profile, then the density
distribution function given in eq.\,(\ref{eq:PDF}) can be written as
\begin{equation}
P(\rho)d\rho = dV/V,
\label{eq:PDFb}
\end{equation}
where $dV/V$ is a fraction of volume with density in the range
$(\rho,\rho+d\rho)$.  The density dependance on radius can be inverted
to give us the radius at a given density $r=r(\rho)$. Then the PDF in
eq.(\ref{eq:PDFb}) takes the form
\begin{equation}
P(\rho) = -\frac{4\pi r^2(\rho)}{V} \frac{dr(\rho)}{d\rho}.
\label{eq:PDFc}
\end{equation}
This can be applied, for example, to the NFW profile. It is
easy to see the trend, if the density profile is a power-law
$\rho\propto r^{-\alpha}$ with the slope  $\alpha$. In this case 
\begin{equation}
\rho^2P(\rho) \propto \rho^{1-3/\alpha}.
\label{eq:PDFd}
\end{equation}
In the outer regions of a dark matter halo the density declines as a
power-law with the slope $\alpha\approx 3$, which implies that
$\rho^2P(\rho)\approx {\it constant}$. It is easy to invert the NFW
profile numerically. Results show that for the halo masses in the
range $M=(10^{12}-10^{15})\Msunh$ the product $\rho^2P(\rho)$ is nearly
constant for overdensities $\rho=10^2-10^4$ and declines at
much larger densities. This decline is consistent with the fact that the
slope $\alpha$ becomes smaller at radii comparable with the
characteristic scale radius $r_s$ of the NFW profile.  Adding results for many
halos with different masses and concentrations will change the
behavior of the $\rho^2P(\rho)$ trend at $\rho >10^4$, but not  for the range
$\rho=10^2-10^4$, where it should remain flat because it is also flat for
each halo.

In summary, both the spherical infall model and the dark matter halo
profiles indicate that the leading term in the density distribution
function should be $P(\rho)\propto\rho^{-2}$. The trend should be
modified by adding suppression on large and small scales. However
insightful, the spherical infall model or the NFW results at large
densities cannot be used to produce accurate results for the density
distribution function. We use those hints to construct our own
approximation for the PDF $P(\rho)$.

Motivated by these results and by our simulations we design our own model. It is nearly a
power-law $P(\rho)\propto\rho^{-\alpha}$ with the slope  $\alpha \approx 2$ that is
truncated with exponents on both the small and large densities. We
call this model {\bf double-exponential}. We tested different
shapes for the exponential terms and find that the following
expression provides errors less than few percent at all redshifts,
smoothing scales, and cosmologies that we study:

\begin{equation}
P(\rho)   = A\rho^{-\alpha}\exp\left[-\left(\frac{\rho_0}{\rho}\right)^{1.1}\right]
                           \exp\left[-\left(\frac{\rho}{\rho_1}\right)^{0.55}\right],
\label{eq:dexp}  
\end{equation}
where $A,\, \alpha$,\, $\rho_0$, and $\rho_1$ are free parameters. As noticed
above the slope $\alpha \approx 2$. The slopes in the exponential
terms 0.55 and 1.1 are results of the fitting of numerical PDFs at
different smoothing scales and redshifts. One may  expect that
adding two more free parameters to the approximation (i.e., the slopes in
the exponential terms) may further improve the quality of the fits. We
find that this is not the case: the data prefer the same slopes
regardless of the value of $\sigma$.

The double-exponential model has four formal free parameters. One may use three
constraints to limit the parameters: the total mass and volume should
be equal to unity (see eq.\,(\ref{eq:normalization})), and the second
moment of the PDF should be equal to $\sigma$ measured in simulations
(see eq.\,(\ref{eq:sigma})). Note that there must be a degree of
freedom left after fixing the constraints otherwise the model would
not be able to reproduce the numerical results that show that the PDF is not
defined solely by $\sigma$, and depends on both the redshift and
$\Omega_m$. The double-exponential model has this additional degree of
freedom.

In practice, we use all four parameters to fit the numerical data. We
typically find that the best-fit parameters provide PDF approximations
that within 1-2\% conserve the mass and match well the numerical value of
$\sigma$ measured in the simulations. The volume is conserved within
1-5\% accuracy.

\makeatletter{}

\section{Simulations}
\label{sec:sim}

Numerical parameters of our simulations are presented in
Table~\ref{table:simtable}, which gives  box
size,  number of particles,  mass of a particle $m_p$,
number of mesh points $N_g^3 $ (if relevant), cell size of the
density/force mesh $\epsilon$, the number of time-steps $N_s$, 
cosmological parameters $\sigma_8$ and $\Omega_m$, and number of
realizations $N_r$.

In order to estimate the density distribution function, we split
each simulation box with a $3-D$ mesh with the cell size $\Delta x$. We
then use the Cloud-In-Cell (CIC) density assignment to generate the
density field. Many filtering sizes $\Delta x$ were used for each
simulation and snapshot.

Different codes were used to make those simulations. The MultiDark
$1~Gpc/h$ simulation (MDPL1) \citep{Klypin2016} was done with the GADGET-2
code \citep{Gadget2}. The ART code \citep{ART} was used to produce the
BolshoiP simulation \citep{Bolshoi}. These two simulations have the
largest resolution and the largest number-density of
particles. However, there are only two of these simulations because
they are very expensive computationally.  Other simulations were carried out with the
parallel Particle-Mesh code GLAM \citep{GLAM}. Because the GLAM code
is much faster, we have many realizations of the same cosmological and
numerical parameters. Simulations B0.5 and D0.5 were designed for
testing possible dependance of the PDF on the matter density
$\Omega_m$. These simulations have the same random seeds to make
comparisons of results easier. 

 All the simulations were started at
initial redshift $z_{\rm init}=100$ using the Zeldovich approximation.
The simulations span three orders of magnitude in mass resolution, a factor of
hundred in force resolution and differ by a factor of 500 in effective
volume (see Table~\ref{table:simtable}). Altogether, we use about 7000
simulations to study the density distribution function. To our
knowledge this is the largest set of simulations available today for this type of analysis.

Analysis of different numerical effects in the estimates of the PDF is 
presented in Appendix~A. In
summary, our results are mostly dominated by systematics, not by
the finite-volume simulation variance. When dealing with individual simulations such as
MDPL1 or BolshoiP we use only bins with more than $N>100$ cells per
bin. For the large sets of simulations A0.5, A1.5, A2.5 we accept bins with
more than 10 cells. The discreteness of density assignment may become
an issue at small densities while the force resolution may affect the
high-density tail of the PDF. We find limits on numerical parameters
that should be satisfied to produce the $P(\rho)$ with errors less than
few percent: (1) the filtering scale $\Delta x$ must be resolved with
not less than 8 force resolution elements: $\Delta x> 8\epsilon$, and
(2) the number of particles per filtering cell should be not less than
10-20.

\makeatletter{}\begin{figure*}
\centering
\includegraphics[width=0.33\textwidth]
{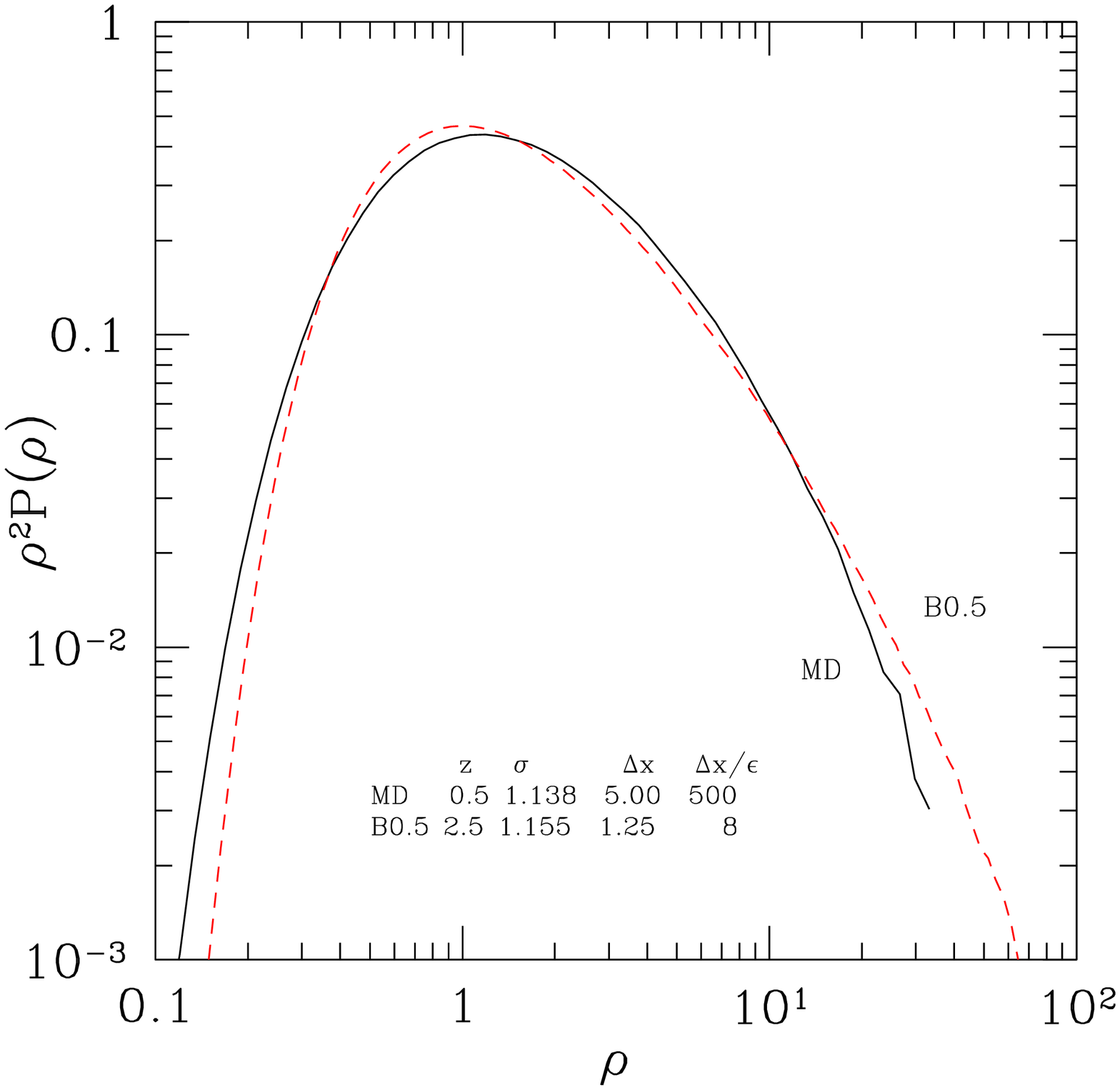}
\includegraphics[width=0.33\textwidth]
{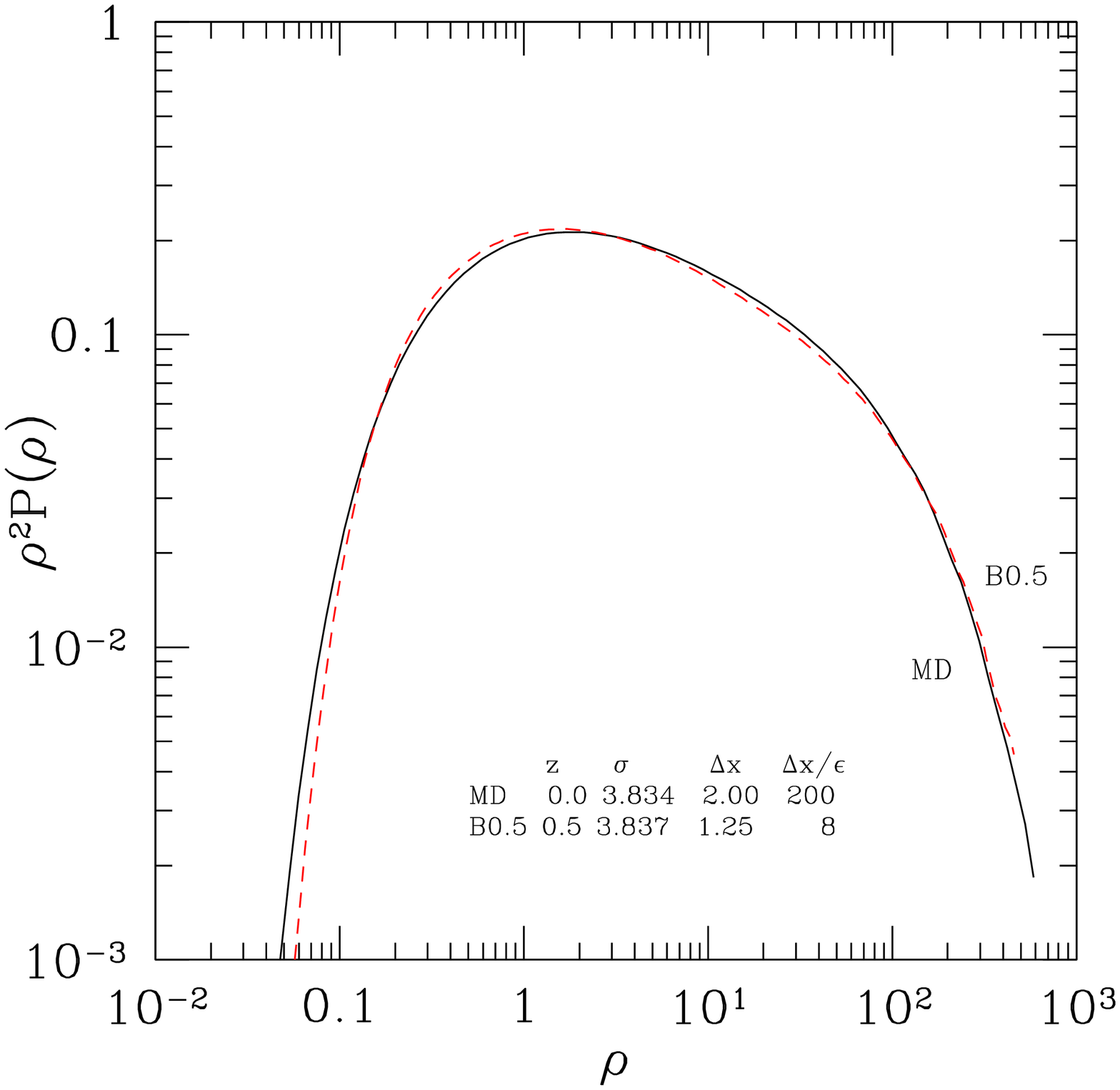}
\includegraphics[width=0.33\textwidth]
{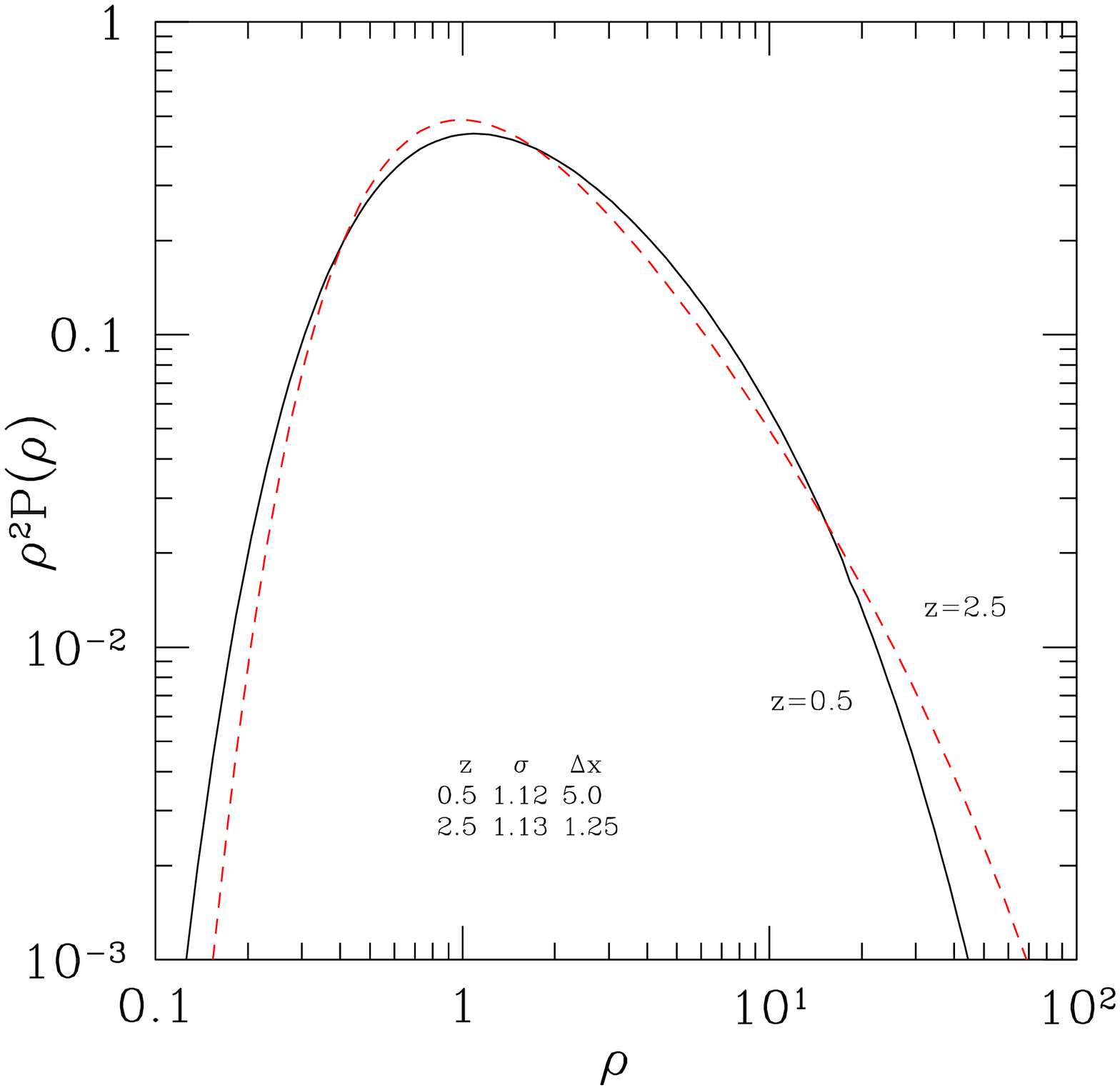}
\caption{Dependance of $P(\rho)$ on redshift at fixed
  $rms$ amplitude of perturbations $\sigma$. Two left panels show examples for
  different redshifts and for different $\sigma$ in the $N$-body
  simulations. In each case we select nearly identical $\sigma$. The
  left panel is for relatively small $\sigma\approx 1$ and for
  $z>0.5$. The middle panel is for high $\sigma$ and $z<0.5$. The
  distribution function clearly depends on redshift though the
  differences are relatively minor for low redshifts. The right panel
  presents analytical estimates based on the spherical infall model
  for the same parameters selected for the left panel. The model
  reproduces the same trend of $P(\rho)$ with redshift. }
\label{fig:amplt}
\end{figure*}

\section{PDF:  main trends}
\label{sec:main}

The overall dependance of $P(\rho)$ on filtering scale and
$rms$ density fluctuation $\sigma$ is illustrated in Figures~\ref{fig:pdf} and \ref{fig:pdf2}.
Full curves in the plots show the results from our
simulations. At small $\sigma$ and large $\Delta x$ the
PDF has a peak at $\rho\approx 1$ that shifts to smaller densities as
$\sigma$ increases (and smoothing scale $\Delta x$ decreases). 
At the same time the distribution function becomes
extremely wide and develops a distinct $\rho^{-2}$ power-law trend that expands
both to large and small densities. By scaling out the $\rho^{-2}$
dependance (Figures~\ref{fig:pdf2}) we reduce the dynamical range and can see better
details of the PDF. In particular, we find a very steep decline
of $\rho^{2}P(\rho)$ on both high and low density tails. The
double-exponential model eq.(\ref{eq:dexp}) was tuned to find
the shape of both declines. Our results for different filtering
scales and different redshifts show that the decline at the large-density
limit is $\propto \exp[-(\rho/\rho_1)^{\nu}]$ and at the small-density
limit it is $\propto \exp[-(\rho/\rho_0)^{-2\nu}]$ with $\nu\approx 0.55$.

Note that at very large $\sigma \gtrsim 10$ the peak of
$\rho^{2}P(\rho)$ has a nearly constant amplitude
$\rho^{2}P(\rho)\approx 0.11$ but the position of the peak shifts to
larger values of $\rho$. Because the total mass must be preserved
($\int \rho P(\rho)d\rho =1$), this implies that at intermediate
scales $\rho\approx 1-100$ the PDF $P(\rho)$ should decline when
$\sigma$ increases, and that the slope $\alpha$ in eq.(\ref{eq:dexp})
must become slightly shallower with increasing $\sigma$.

It is often taken for granted that the PDF depends only on the
amplitude of the density perturbations on a given filtering scale
$\sigma$. Indeed, this is the dominant behavior of $P(\rho)$. However,
this is not exactly correct. Our simulations have such a good accuracy
that now we can test the dependance of the PDF on redshift at
fixed $\sigma$ and on cosmological parameters.

We first select redshifts and smoothing scales in such a way that
$\sigma$ for two different redshifts are nearly
identical. Figure~\ref{fig:amplt} presents two examples of such cases
-- one for relatively low $\sigma\approx 1$ and another for larger
$\sigma\approx 4$. The differences between PDFs at the same $\sigma$
and different $z$ are not large: 5-10\% depending on the density
where the differences are measured. Nevertheless, the differences
clearly exist. We also estimate the differences using the spherical
infall model and present the results in the right panel of
Figure~\ref{fig:amplt}.

\makeatletter{}\begin{figure*}
\centering
\includegraphics[width=0.45\textwidth]
{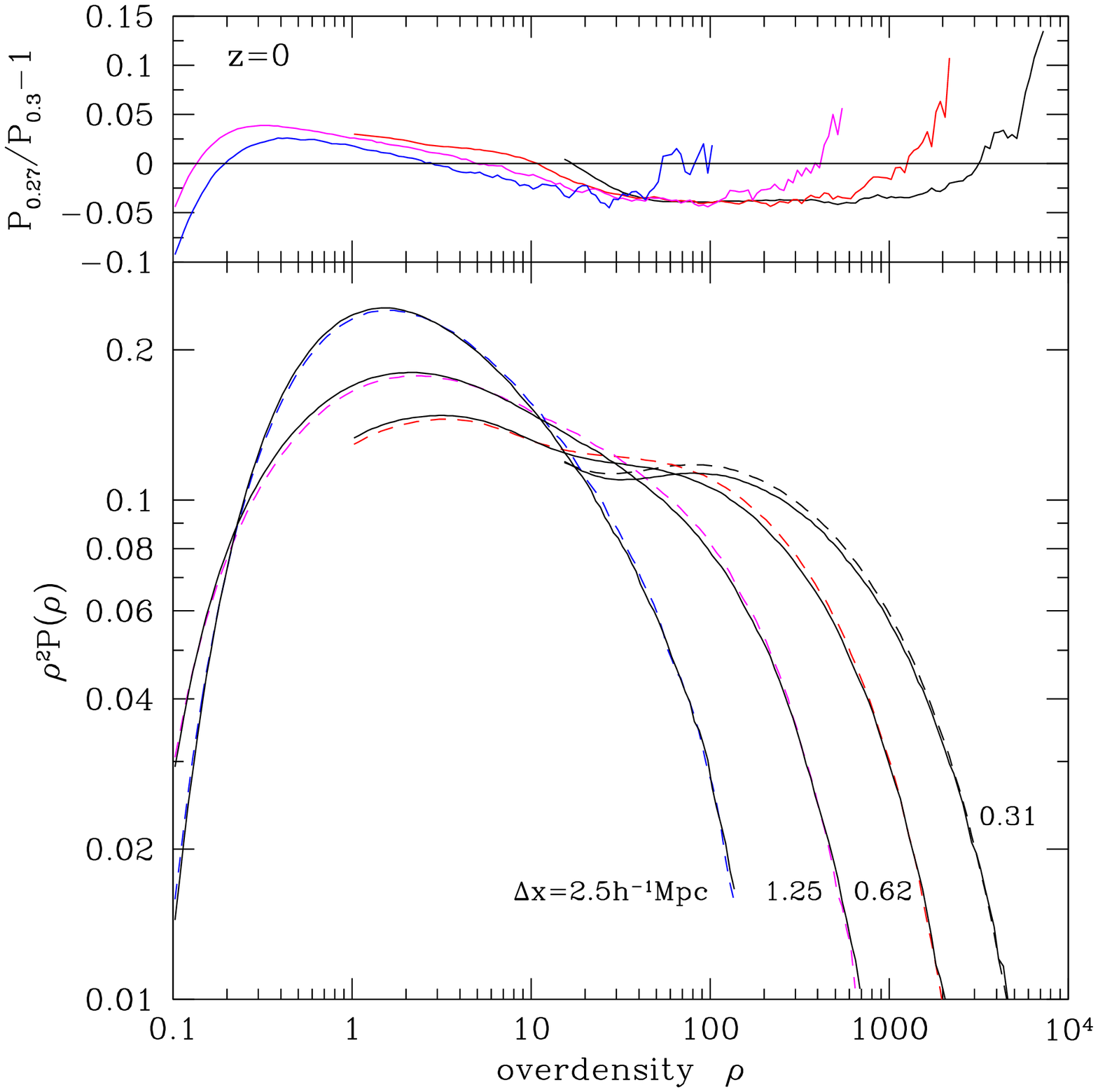}
\includegraphics[width=0.45\textwidth]
{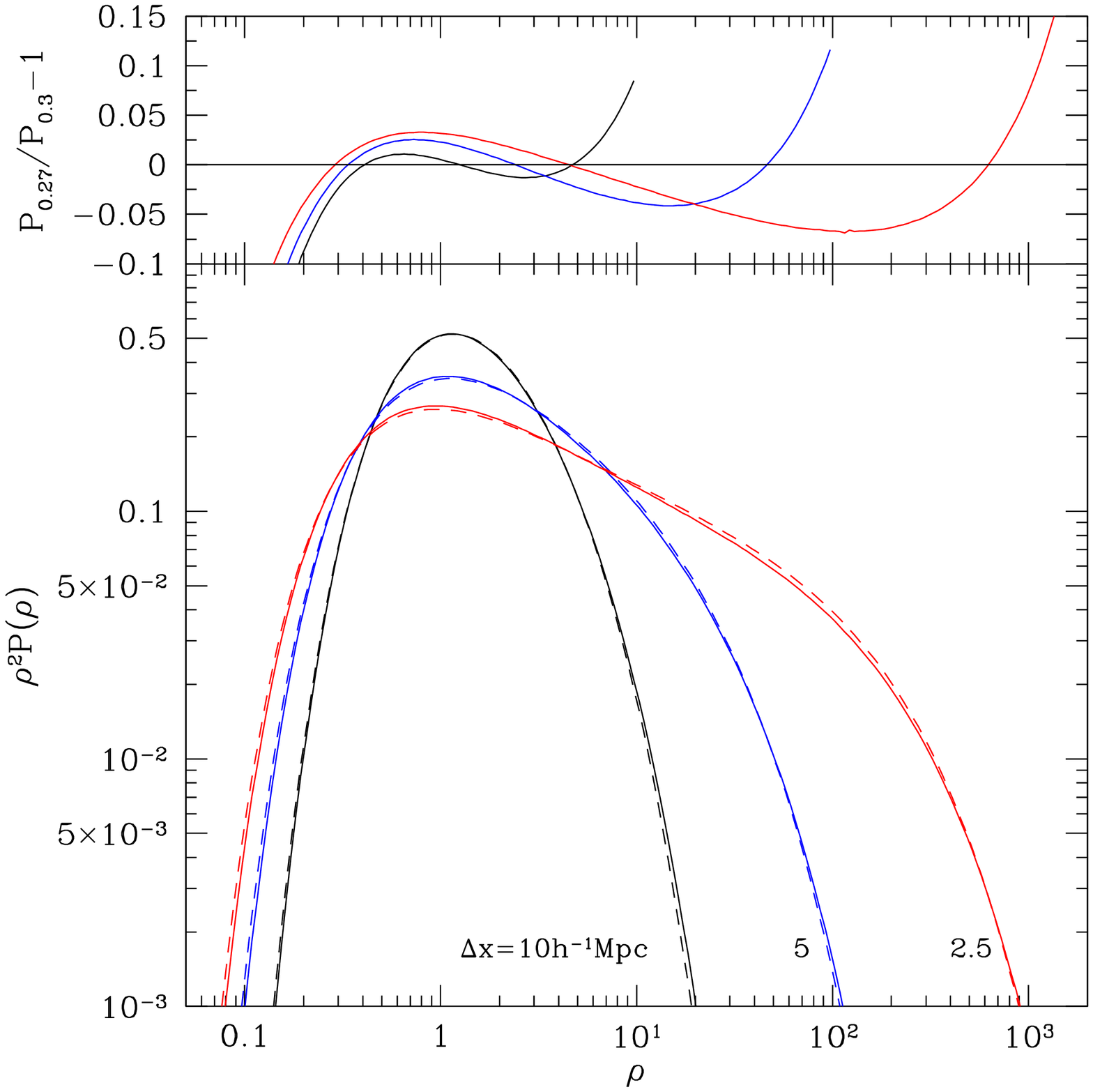}
\caption{Dependance of $P(\rho)$ on the cosmological matter density
  $\Omega_m$. Dashed (full) curves show results for models with
  $\Omega_m=0.307$ ($\Omega_m=0.270$) at $z=0$ for different filtering
  scales. Other cosmological parameters are the same for both
  models. The left panel presents comparison of $N$-body simulations
  B0.5 and D0.5. The right panel shows results of the spherical infall
  model. The density distribution function weakly but systematically
  depends on $\Omega_m$ with $\sim 5$\% deviations at different
  densities.}
\label{fig:Omega}
\end{figure*}

\makeatletter{}\begin{figure*} \centering
\includegraphics[width=0.45\textwidth]
{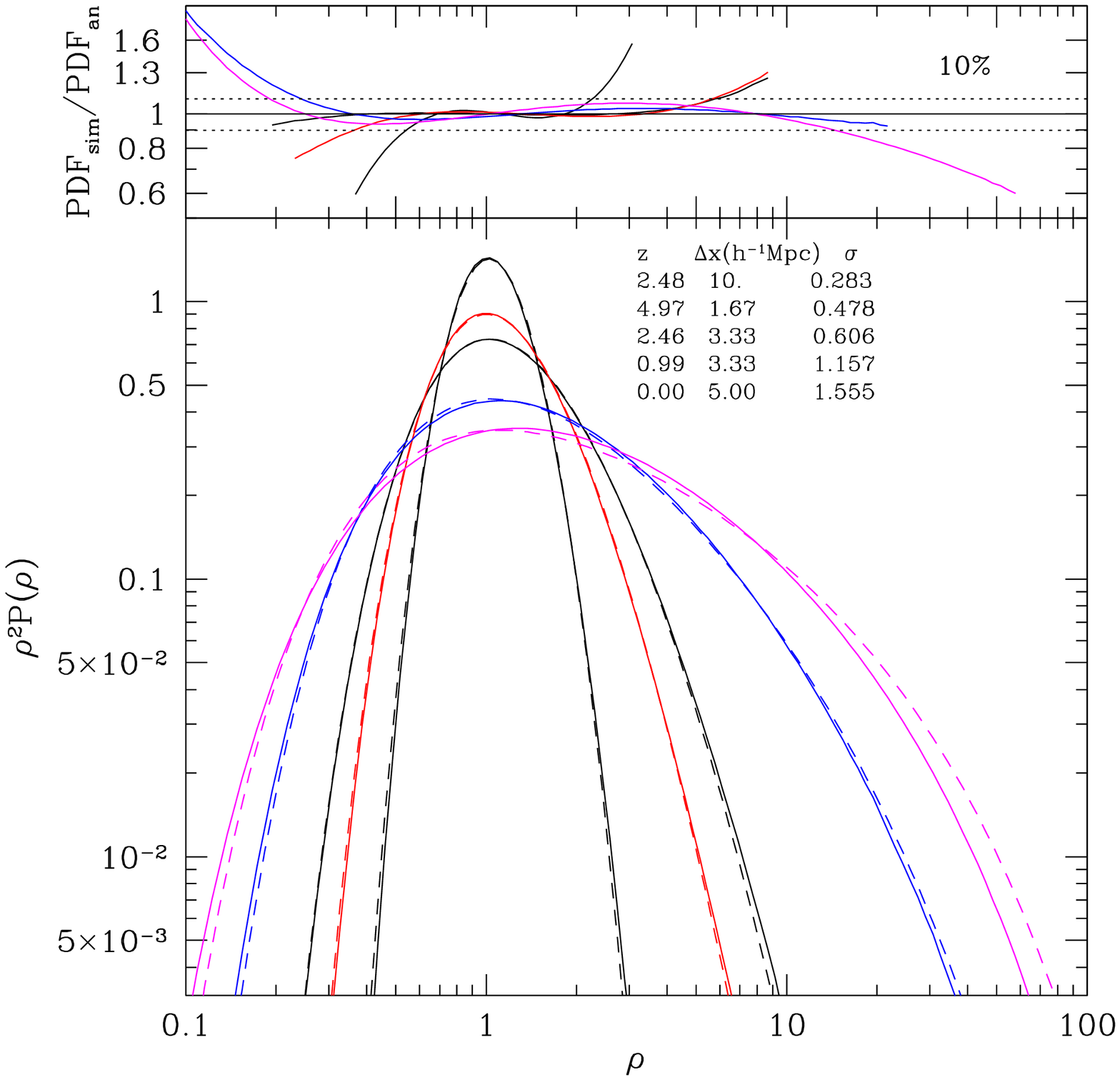}
\includegraphics[width=0.45\textwidth]
{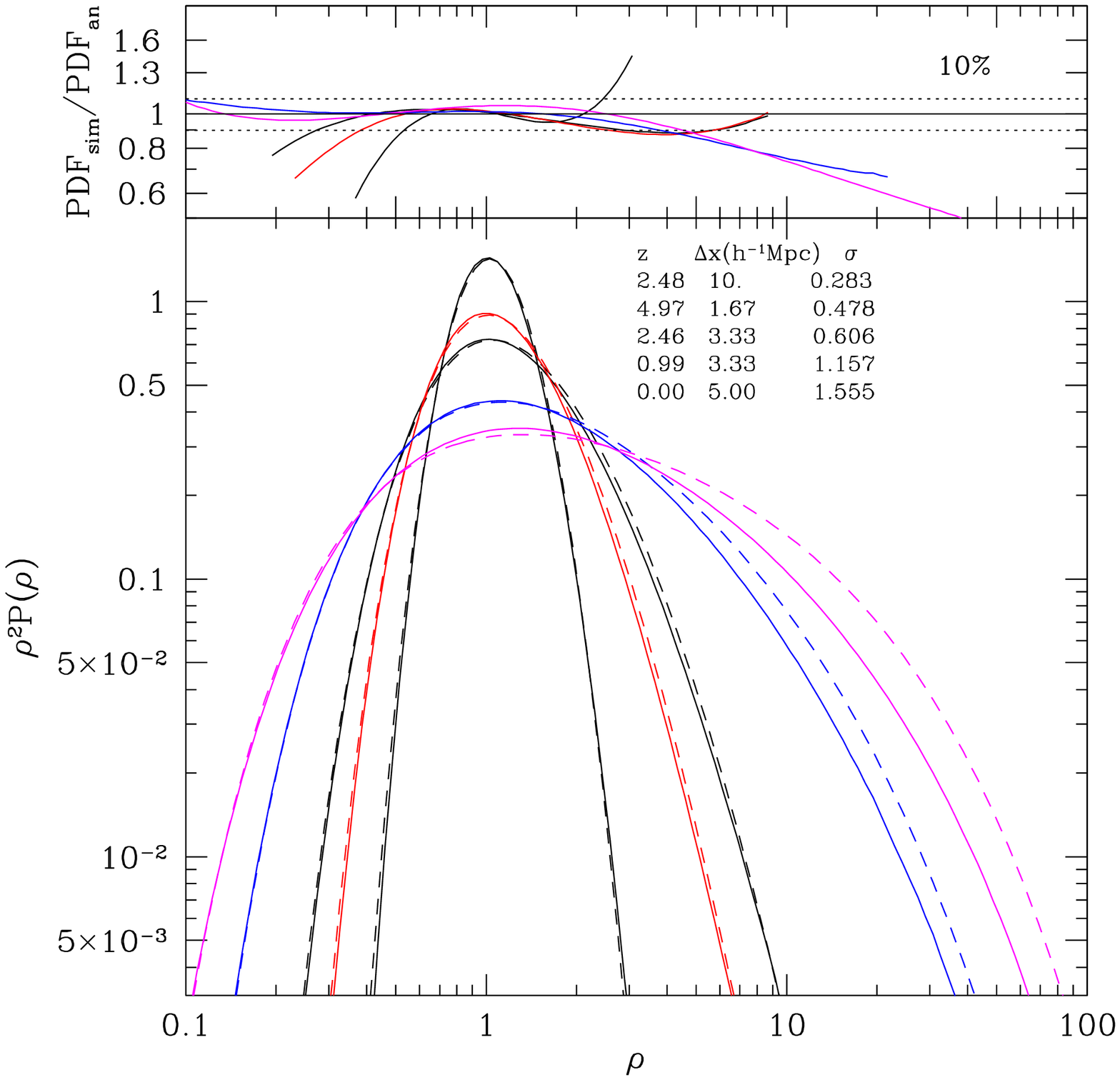}
\caption{Accuracy of spherical infall model in the regime of
  $\sigma\lesssim 1$. The left panel presents results for the
  approximation eq.\,(\ref{eq:sph1}). The right panel is for
  eq.\,(\ref{eq:sph2}). We select different redshifts and different
  filtering scales. Full curves in the bottom panels show results of
  simulations while the dashed curves are for the analytical
  approximation. The top panels present relative errors of the
  approximations.  }
\label{fig:spher}
\end{figure*}

The PDF also slightly depends on parameters of the cosmological
model. In the left panel of Figure~\ref{fig:Omega} we compare $z=0$
results for the B0.5Gpc and D0.5Gpc simulations that differ only by the
matter density parameter $\Omega_m$. Again, differences are small but
clearly present at $\sim 5-10\%$ level. The right panel shows
predictions for the spherical infall model with the same basic
conclusion: PDF does depend on $\Omega_m$.

One can understand why the PDF depends on $z$ and $\Omega_m$, if one
realizes that at any given density $\rho$ the value of $P(\rho)$ is
formally a functional on the non-linear power spectrum. This means
that it depends not only on $\sigma(R)$ but also on the whole shape of
the power spectrum.  We can analyze the situation by assuming that
$P(\rho)$ depends just on $\sigma_{L}(R)$ (see eq.\,(\ref{eq:sigmaL})) and on its local logarithmic
derivative $\gamma$ at scale $R$ as defined by
eq.\,(\ref{eq:gam}). Then the dependence of the PDF on redshift for
a fixed value of $\sigma_{L}(R)$ can  be explained, because at
larger redshits the given value of $\sigma_{L}$ is attained at a
smaller scale $R$, where $\sigma_{L}(R)$ is less steep for a CDM-type
power spectrum.

If we look at different terms for $\rho^2P(\rho)$ in
eq.\,(\ref{eq:sph1}), then we note that the $\sigma_L$ in the argument
of the exponential term is also a function of $\rho$ (falling faster
with $\rho$ as $\gamma$ grows). So, for larger values of $\gamma$ the
PDF will be smaller for large $\rho$ values, where the exponential
behavior dominates, while for small values of $\rho$ it will be
larger. The derivative in the right-hand-side of eq.\,(\ref{eq:sph1})
also depends on $\gamma$: it will be larger for larger $\gamma$ values.
However, it is only important for the intermediate values of density, where
the behavior can qualitatively be explained by the behavior in the
extremes and the conservation of probability. In short, 
this all  implies a smaller PDF in the low $\rho$
limit and a larger one in the large $\rho$ limit as compared to the PDF
at smaller redshift. 

The same qualitative behavior would be observed
for smaller values of $\Omega_m$  when studying the dependence of the
PDF on $\Omega_m$  at a given redshift: as $\Omega_m$
decreases the power spectrum flattens (up to scales of the order of
the horizon at the start of matter domination), leading to a less steep
$\sigma_L.$

  \makeatletter{}\begin{figure*} \centering
\includegraphics[width=0.45\textwidth]
{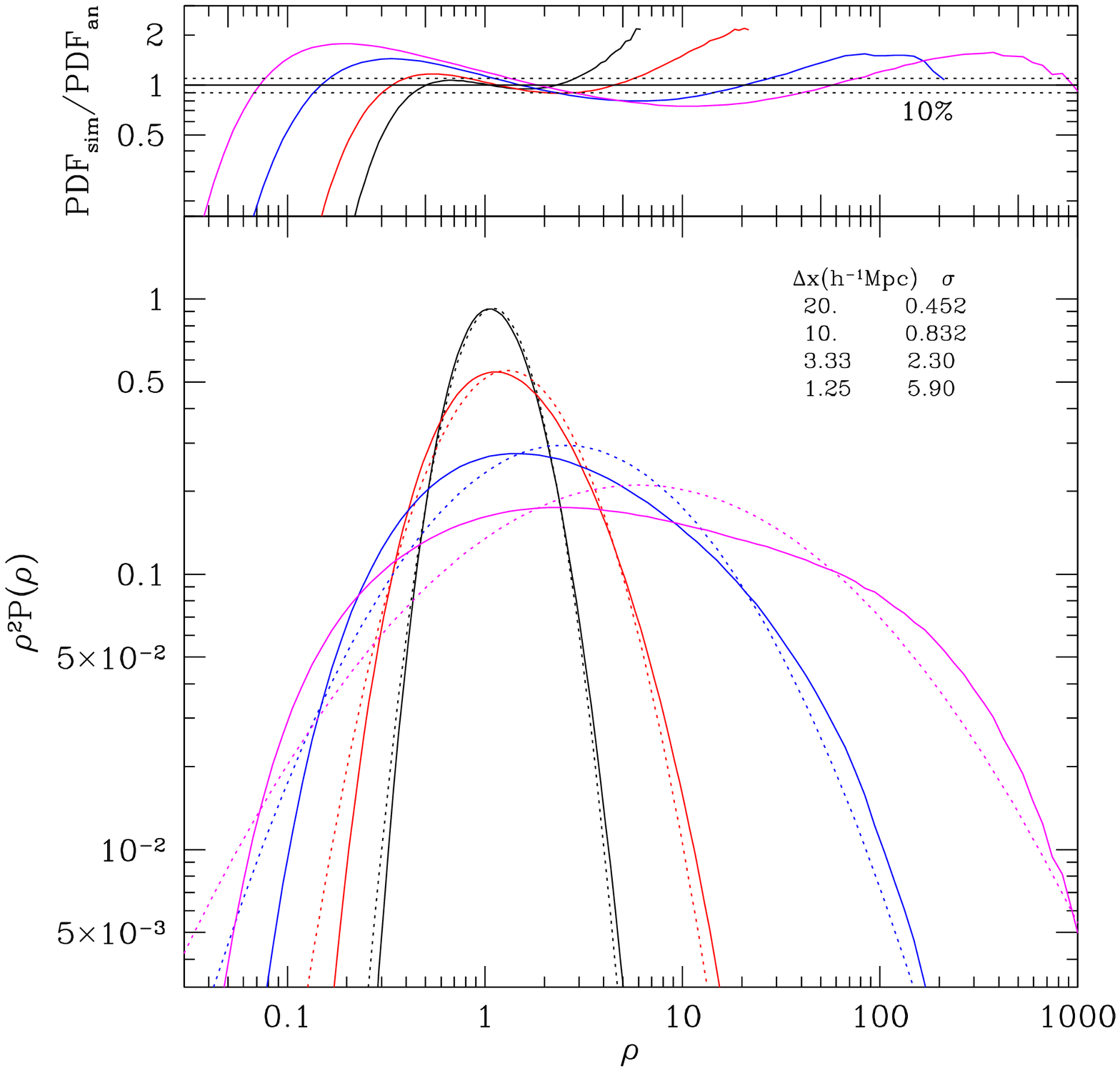}
\includegraphics[width=0.45\textwidth]
{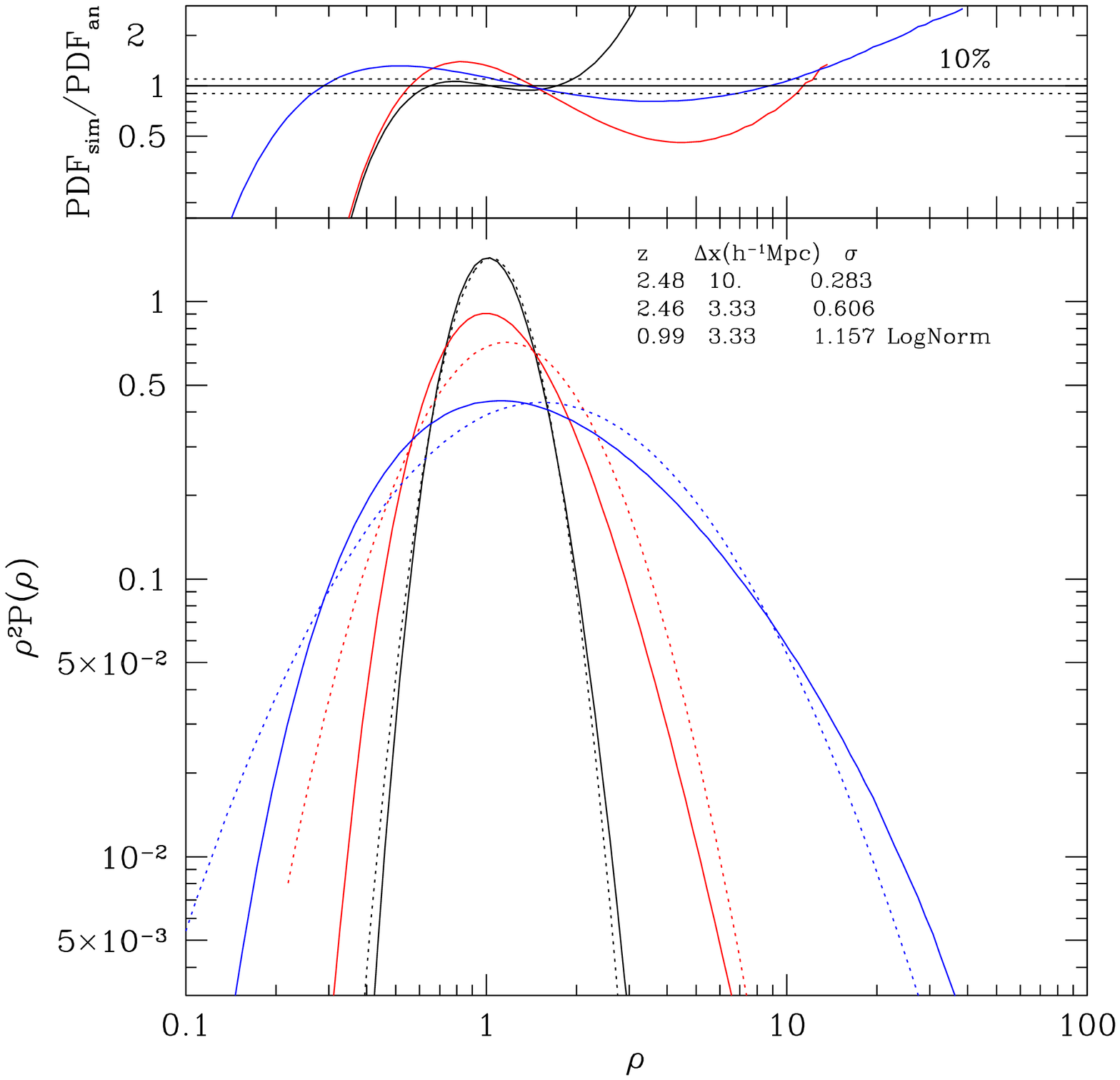}
\caption{Accuracy of the log-normal distribution. {\it Left:} Results
  for different smoothing scales at $z=0$. {\it Right:} Results for
  different redshifts. The log-normal distribution provides accurate
  fits (less than 10\% errors) only for a very limited range of
  densities and $\sigma$. Comparison with the predictions of the spherical infall model
  in Figure~\ref{fig:spher} shows that the log-normal fits for
  $\sigma<1$ always give errors 2-3 times larger than the spherical
  infall model. To make things worse, the log-normal distribution has
  a wrong shape. It predicts wrong position of the maximum; slopes on
  both ends of the PDF are also incorrect. The only advantages of the
  log-normal fits are that it is very simple and that it never fails
  catastrophically. }
\label{fig:lognorm}
\end{figure*}

\makeatletter{}\begin{figure*}
\centering
\includegraphics[width=0.45\textwidth]
{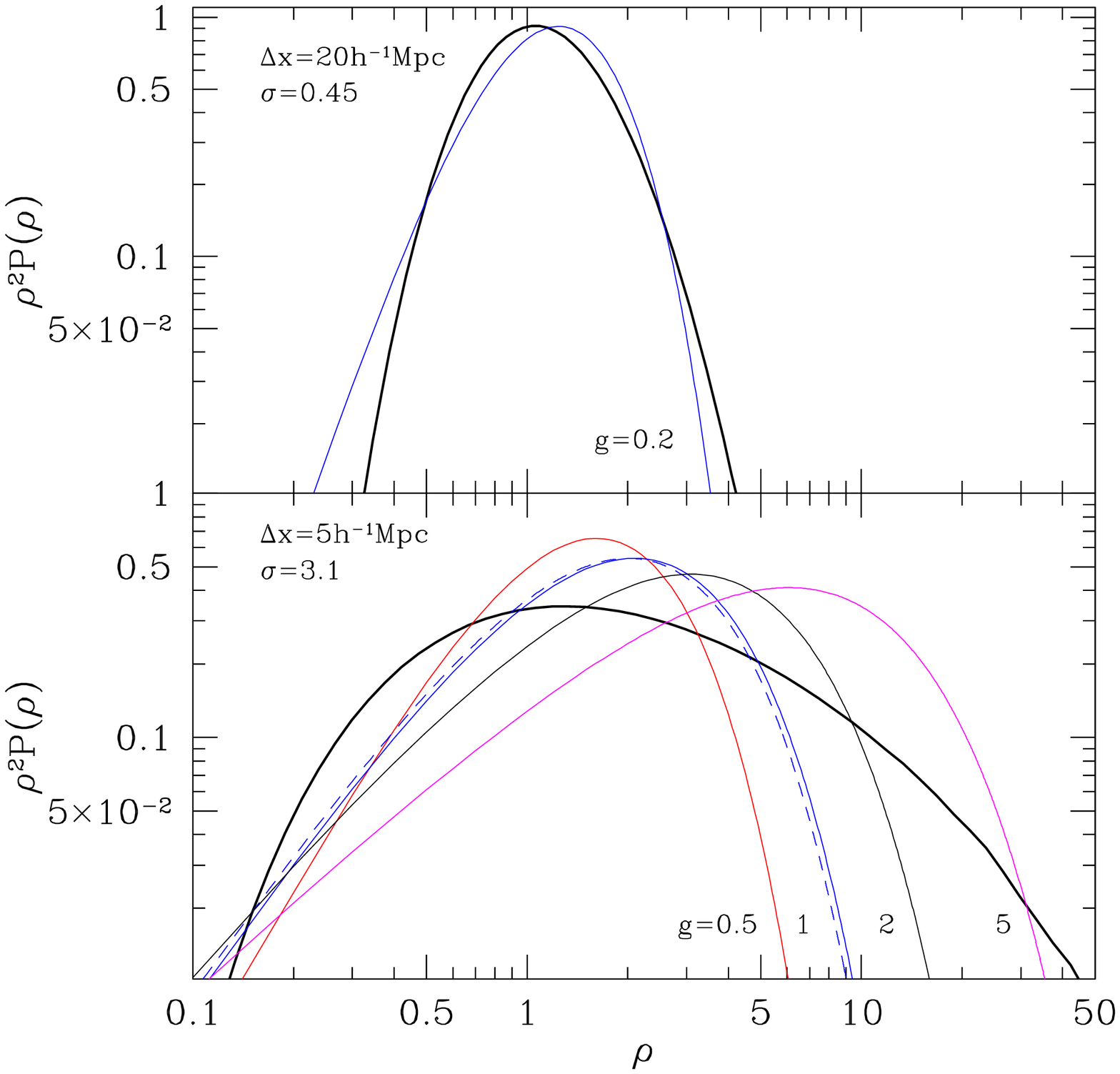}
\includegraphics[width=0.45\textwidth]
{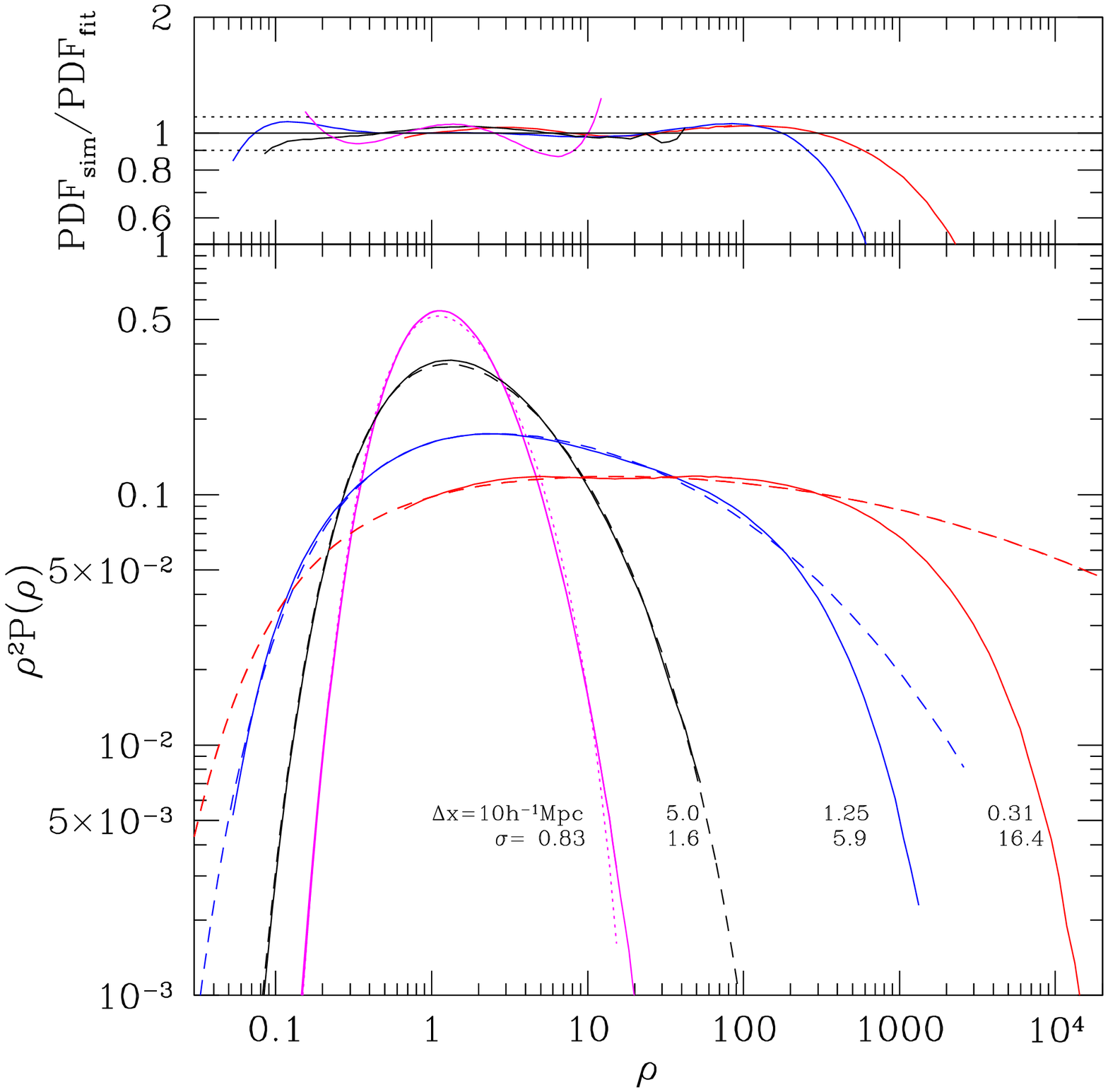}
\caption{Accuracy of different analytical approximations. PDF of
  $N$-body simulations are taken at $z=0$. {\it Left:} Negative
  BiModal (NBN) fits (eq.~(\ref{eq:nbn}) are shown as thin cuves
  labeled by values of $g$ parameter used to make the fits.  $N$-body
  results are the thick lines. NBN provides a reasonable fit
  ($\sim 10\%$ accuracy for $\rho =0.4-3$) for large cell sizes
  corresponding to low $rms$ fluctuations $\sigma<0.5$. The bottom
  panel shows an example of NBN fits for large $\sigma$. It does not
  provide a good fit regardless of what value of $g$ is used. {\it
    Right:} Results of fitting with the generalized extreme value
  (GEV) distribution. Full cuves in the bottom panel show $N$-body
  results; GEV fits are the dashed curves. GEV provides much more
  accurate fits as compared with the log-normal or NBN
  approximations. However, it fails at very large densities for small
  cell sizes.}
\label{fig:NBN}
\end{figure*}

\makeatletter{}\begin{figure*}
\centering
\includegraphics[width=0.45\textwidth]
{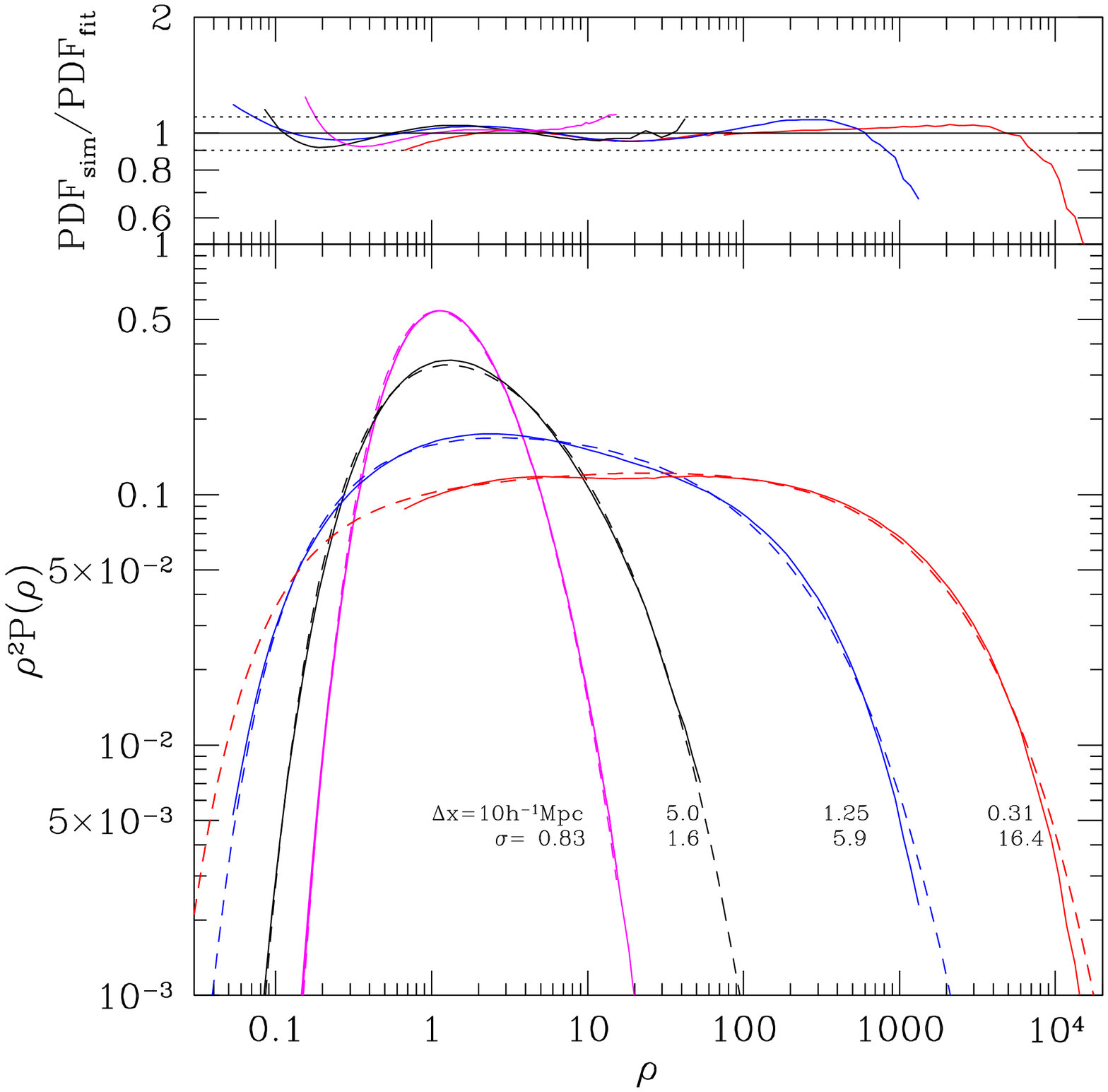}
\includegraphics[width=0.45\textwidth]
{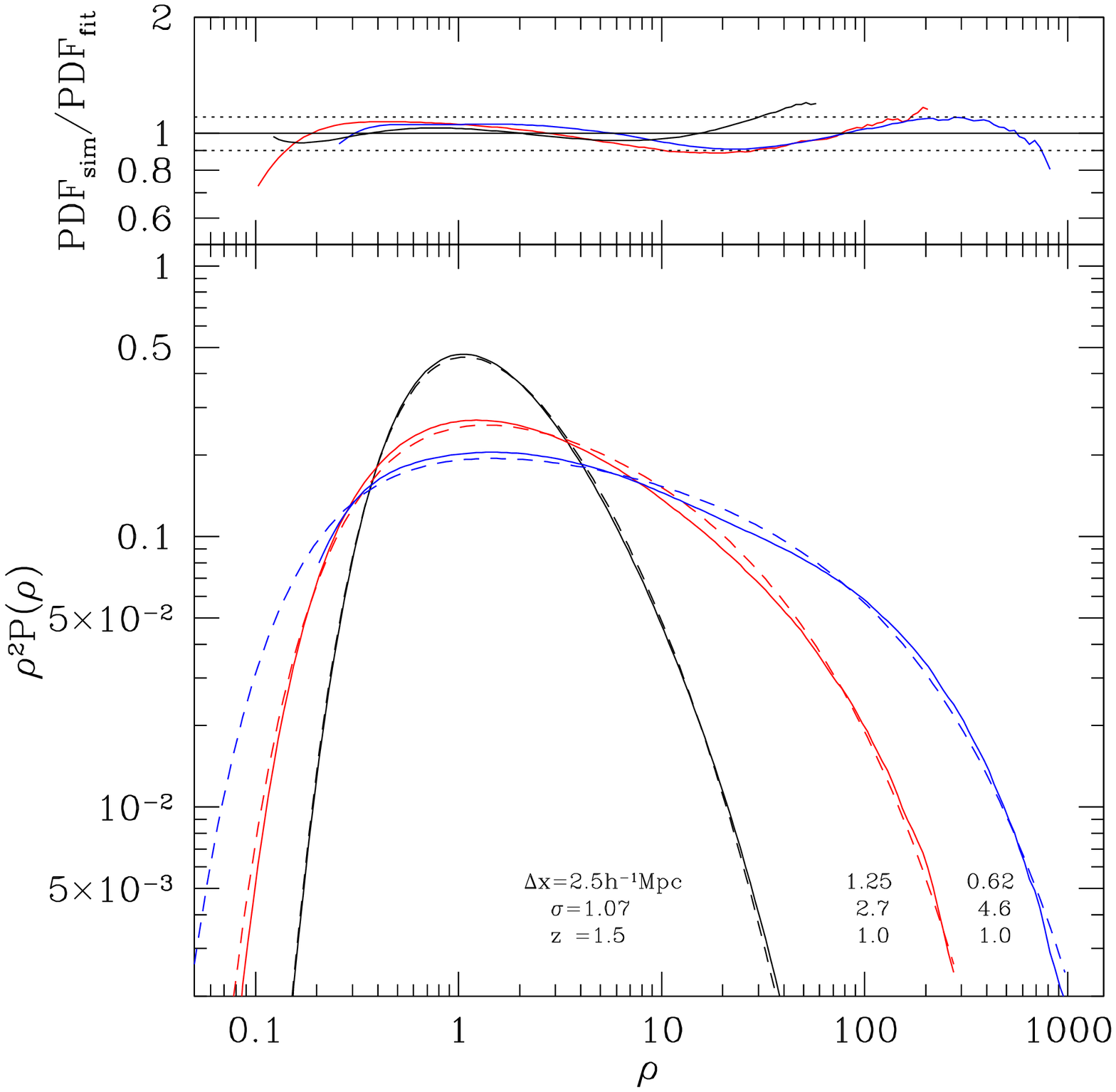}
\caption{Accuracy of the double-exponential model
  given by eq.~(\ref{eq:dexp}). The left panels show results for $z=0$. Results
  of fits for different redshifts are presented in the right panel.}
\label{fig:doubleexp}
\end{figure*}
\makeatletter{}\section{PDF: Testing different approximations and models.}
\label{sec:sphLog}

We start our analysis of different approximations by testing the
spherical infall model and the log-normal distribution. Both models
are expected to work and typically used for relatively low rms
fluctuations $\sigma\lesssim 1$. 

Figure~\ref{fig:spher} shows results for two modifications of the
spherical infall model. We select five configurations with different
filtering scales, $\sigma$ and redshifts. There are some differences
between the pure spherical infall model in eq.\,(\ref{eq:sph1}) and the
excursion set model in eq.\,(\ref{eq:sph2}). For example, the excursion
set model produces smaller errors at $\rho<1$ and $\sigma\approx 1$.
At the same time, it makes visibly larger errors at $\rho>1$. For this
reason we prefer the standard spherical infall model. It provides
smaller than 10\% errors for points that are larger than 0.1 of the
maximum of the PDF. The error increases substantially in the peripheral
regions. If better accuracy is required for the small values of the
PDF, one would need to use results of $N$-body simulations, not the
approximations.

Results for the log-normal distribution are shown in
Figure~\ref{fig:lognorm}, where in the left panel we present $z=0$
results, and results for different redshifts are shown in the right
panel. It is clear that the log-normal distribution produces much
worse fits as compared with the spherical infall model. For example,
errors in the central region are less than 10\% only for
$\sigma<0.5$. They become dramatically worse for even slightly larger
$\sigma$.  For $\sigma>1$ the lognormal distribution makes typically
$\sim 50\%$ errors and, which is even worse, predicts a wrong shape of
the PDF. It predicts a wrong position of the maximum; slopes of the
declining PDF are not correct.

The log-normal distribution has two advantages as compared with the
spherical infall model: (a) It is simple; does not require the
machinery of handling the power spectrum and numerical
derivatives. (b) it makes mediocre predictions that do not totally
fail. We definitely recommend it for quick-and-dirty applications, but
not for accurate estimates.

Left panels in Figure~\ref{fig:NBN} show results for the Negative
Binomial distribution given by eq.\,(\ref{eq:nbn}). For the large smoothing
scale $\Delta x=20\Mpch$ it provides a reasonable fit with errors
$\sim 10\%$ for densities $\rho =0.4-2.5$. However, it is clear that
it has a wrong shape: too steep at large densities and too shallow at
small densities. This becomes a serious issue for large values of
$\sigma$. For example, we could not find any good fit for
$\Delta x=5\Mpch$ shown in the bottom panel. It may be not fare to use
the NBN for the dark matter PDF because we are in a regime that was
not favorable for the NBN: study configurations with very large number
of particles per cell while the NBN was designed to handle very small
number $\bar N$.

The GEV approximation eq.\,(\ref{eq:GEV}) scored much better, as
illustrated in the right panels of Figure~\ref{fig:NBN}. Indeed, it
provided excellent fits for $\sigma \lesssim 2$ with errors less than
10\% and even for larger $\sigma$ it gives very good accuracy, but it
starts to fail catastrophically at very large
densities. \citet{Lee2017} tested this approximation for PDF for the
Bolshoi simulation. Their results are compatible with ours. However,
\citet{Lee2017} seems did not pay attention to the situation at large
densities, where the GEV becomes unacceptable. It still can be useful
for low densities, but the problem is that there is no a priory
estimate at what density and $\sigma$ the GEV fails.

Comparison of the double-exponential model with the $N$-body results
has been already presented in Figure~\ref{fig:pdf}. More detailed
analysis is shown in Figure~\ref{fig:doubleexp}. Parameters of the
fits for different smoothing scales and redshifts are given in 
Appendix~B. Of all approximations studied in this paper the double
exponential model is by far the best. It provides very accurate (few
percent) fits for $\sigma \gtrsim 1$ with densities
$\rho\approx 10^{-1}-10^5$. With somewhat larger errors it still works
down to $\sigma \approx 0.5$. To some degree the success of the
approximation is not surprising because it was designed to reproduce
the main features of the spherical infall model and the PDF of
individual NFW halo profiles that predict the $P(\rho)\propto \rho^{-2}$
trend for large densities. And, of course, the parameters of the
approximation -- the two power-law slopes -- were tuned to produce
best fits.

However, it is unexpected that the exponential terms in
eq.\,(\ref{eq:dexp}) should have the same power-law slopes 0.55 and
1.11 for all densities and smoothing scales. Indeed, we tried
different combinations of the slopes -- even with changing values of
the slopes for different $\sigma$ -- and did not find them to improve
the fits. There is one problem with the constant slopes, though: the
approximation cannot work for a very small $\sigma$ where the PDF must
become a Gaussian.  The eq.\,(\ref{eq:dexp}) does not allow a
transition to a Gaussian distribution. This is not a serious issue
because at $\sigma \lesssim 1$ the spherical infall model provides an
adequate approximation for the dark matter density distribution function.

\makeatletter{}\section{Conclusions}
\label{sec:concl}

Using a large suite of cosmological $N$-body simulations we study the
shape and the evolution of the dark matter one-point probability distribution
function. Unlike most of other studies, we cover a very large range of
smoothing scales $R=100\,\kpch - 20\,\Mpch$ and $rms$ density fluctuations
$\sigma$. We find that as $\sigma$ increases, the PDF becomes a
power-law $P\propto\rho^{-2}$ that is truncated with exponential terms on both small and
large densities. This trend is consistent with the extrapolation of
both the spherical infall model and the PDF expected at high densities
for the NFW density profile of dark matter halos. 

The PDF weakly depends on redshift (at fixed $\sigma$) and on
matter density $\Omega_m$. The effect is relatively small
($\sim 5-10\%$) but is clearly observed in simulations. The spherical
infall model also has the same trend. This behavior contradicts
analytical approximations such as the log-normal or the GEV that assume that the
PDF should depend only on $\sigma$.

The basic trend $P\propto\rho^{-2}$ gives us a motivation to construct
a new model given in  eq.\,(\ref{eq:dexp}), which we call
double-exponential distribution. It formally has four free parameters of which two
can be fixed by requiring that the total volume and mass must be equal
to unity. The model works only for $\sigma\gtrsim 1$ and does not
allow the transition to a  Gaussian distribution as expected for
$\sigma\ll 1$. Nevertheless, for $\sigma\gtrsim 1$ the model gives the
best performance of all approximations that we tested in this work, with errors of just
few percent for $P(\rho)$ when the PDF changes by 12 orders of
magnitude. Parameters of the double-exponential model are provided in
Appendix~B. The model potentially may be modified by adding few extra
parameters to allow for accurate treatment at small $\sigma$. We did
not try to do it for two reasons: (1) the spherical infall model gives
accurate enough treatment for this regime, and (2) it is cheap to make
adequately accurate $N$-body simulations for $\sigma<1$ if needed.

The spherical infall model provides accurate predictions for the
$N$-body PDF results for low values of $\sigma<1$, but it becomes less reliable for larger
$\sigma$, which is expected for this model. The combination of the
double-exponential model at large $rms$ fluctuation with the spherical
infall model in the small $rms$ regime yields a remarkably accurate
density distribution for all regimes of clustering of the cosmological
matter field.

We also tested different analytical approximations. The often used
log-normal distribution, as was shown before, does not make good fits
and needs significant modifications before it can provide accurate
results. It is pretty much useless for large densities because it does
not provide a path to explain the main trend at large densities,
i.e. the power-law trend $P\propto\rho^{-2}$. It clearly has an
advantage of being simple, and it does not fail catastrophically as shown in
Figure~\ref{fig:lognorm}. Our results for $\sigma<1$ demonstrate that the
log-normal approximation always made significantly worse fits as
compared with the spherical infall model predictions. In addition, the log-normal
approximation cannot accomodate the dependance of the PDF on
redshift and $\Omega_m$, while the spherical infall model nicely
reproduces the effect.

The GEV approximation scores much better than the log-normal distribution. Even for
a large $\sigma\approx 10$ it gives remarkably accurate results for
densities up to $\rho\approx 10^3$. However, the approximation fails
catastrophically at larger densities, and there is no obvious way to
predict at what density it still works or fails.

While it is useful and insightful to have analytical models for the
PDF, one does not really need them if $\sim 1$\% accuracy is a
requirement and $\sigma\gtrsim 0.5$.  Then there is no alternative to
$N$-body simulations: one can get very accurate, fast and cheap
results. The cost of one GLAM A0.5 simulation is just 3.5~hrs of
wall clock on a modest data server \citep{GLAM}. For the B0.5 run it
is 10~hrs. In this paper we analyse thousands of realizations, but this was to
prove that one does not need those to make accurate PDF: just a few
realizations is enough. However, one needs to be carefull about
numerical effects when using these fast Particle-Mesh
simulations. \citet{GLAM} provide detailed description of constraints
for the simulations, and Appendix~A in this paper gives prescriptions
on how to use the $N$-body simulations to reach $\sim 1\%$ accuracy in
the PDF.

\section*{Acknowledgements}

A.K. acknowledges support of the Fulbright Foundation and support of
the Instituto de Fisica Teorica, CSIC, Madrid,
Spain. F.P. acknowledges support from the Spanish MINECO grant
AYA2014-60641-C2-1-P.  FDA acknowledges financial support from ``la
Caixa''-Severo Ochoa doctoral fellowship. We thank Johan Comparat (IFT,
Madrid), Claudia Scoccola (IAC, Tenerife), and Sergio Rodriguez-Torres
(UAM, Madrid) for comments and fruitful discussions.  The PPM-GLAM
simulations have been performed on the FinisTarrae II supercomputer at
CESGA in Galicia, supported by the Xunta de Galicia, CISIC, MINECO and
EU-ERDF.

\bibliography{PDF}

\begin{thebibliography}{}

\bibitem[\protect\citeauthoryear{{Bel} et~al.,}{{Bel} et~al.}{2016}]{Bel2016}
{Bel} J.  et~al., 2016, \aap, 588, A51

\bibitem[\protect\citeauthoryear{{Bernardeau}}{{Bernardeau}}{1994}]{Bernardeau%
1994}
{Bernardeau} F.,  1994, \aap, 291, 697

\bibitem[\protect\citeauthoryear{{Betancort-Rijo}}{{Betancort-Rijo}}{1991}]{Ju%
an1991}
{Betancort-Rijo} J.,  1991, \mnras, 251, 399

\bibitem[\protect\citeauthoryear{{Betancort-Rijo}}{{Betancort-Rijo}}{2000}]{Ju%
anNB}
{Betancort-Rijo} J.,  2000, Journal of Statistical Physics, 98, 3

\bibitem[\protect\citeauthoryear{{Betancort-Rijo} \&
  {L{\'o}pez-Corredoira}}{{Betancort-Rijo} \&
  {L{\'o}pez-Corredoira}}{2002}]{Juan2002}
{Betancort-Rijo} J.,  {L{\'o}pez-Corredoira} M.,  2002, \apj, 566, 623

\bibitem[\protect\citeauthoryear{{Carron}, {Wolk} \& {Szapudi}}{{Carron}
  et~al.}{2015}]{Carron2015}
{Carron} J.,  {Wolk} M.,    {Szapudi} I.,  2015, \mnras, 453, 450

\bibitem[\protect\citeauthoryear{{Chuang} et~al.,}{{Chuang}
  et~al.}{2015}]{Chuang2015}
{Chuang} C.-H.  et~al., 2015, \mnras, 452, 686

\bibitem[\protect\citeauthoryear{{Clerkin} et~al.,}{{Clerkin}
  et~al.}{2017}]{Clerkin2017}
{Clerkin} L.  et~al., 2017, \mnras, 466, 1444

\bibitem[\protect\citeauthoryear{{Coles} \& {Jones}}{{Coles} \&
  {Jones}}{1991}]{ColesJones}
{Coles} P.,  {Jones} B.,  1991, \mnras, 248, 1

\bibitem[\protect\citeauthoryear{{Colombi}}{{Colombi}}{1994}]{Colombi1994}
{Colombi} S.,  1994, \apj, 435, 536

\bibitem[\protect\citeauthoryear{{Efstathiou}, {Kaiser}, {Saunders},
  {Lawrence}, {Rowan-Robinson}, {Ellis} \& {Frenk}}{{Efstathiou}
  et~al.}{1990}]{Efstathiou1990}
{Efstathiou} G.,  {Kaiser} N.,  {Saunders} W.,  {Lawrence} A.,
  {Rowan-Robinson} M.,  {Ellis} R.~S.,    {Frenk} C.~S.,  1990, \mnras, 247,
  10P

\bibitem[\protect\citeauthoryear{{Gazta{\~n}aga}, {Fosalba} \&
  {Elizalde}}{{Gazta{\~n}aga} et~al.}{2000}]{NegBinom2000}
{Gazta{\~n}aga} E.,  {Fosalba} P.,    {Elizalde} E.,  2000, \apj, 539, 522

\bibitem[\protect\citeauthoryear{{Hamilton}}{{Hamilton}}{1985}]{Hamilton1985}
{Hamilton} A.~J.~S.,  1985, \apjl, 292, L35

\bibitem[\protect\citeauthoryear{{Hubble}}{{Hubble}}{1934}]{Hubble1934}
{Hubble} E.,  1934, \apj, 79, 8

\bibitem[\protect\citeauthoryear{{Hurtado-Gil}, {Mart{\'{\i}}nez},
  {Arnalte-Mur}, {Pons-Border{\'{\i}}a}, {Pareja-Flores} \&
  {Paredes}}{{Hurtado-Gil} et~al.}{2017}]{Hurtado2017}
{Hurtado-Gil} L.,  {Mart{\'{\i}}nez} V.~J.,  {Arnalte-Mur} P.,
  {Pons-Border{\'{\i}}a} M.~J.,  {Pareja-Flores} C.,    {Paredes} S.,  2017,
  ArXiv e-prints

\bibitem[\protect\citeauthoryear{{Kitaura} et~al.,}{{Kitaura}
  et~al.}{2016}]{Kitaura2016}
{Kitaura} F.-S.  et~al., 2016, \mnras, 456, 4156

\bibitem[\protect\citeauthoryear{{Klypin} \& {Prada}}{{Klypin} \&
  {Prada}}{2017}]{GLAM}
{Klypin} A.,  {Prada} F.,  2017, ArXiv e-prints

\bibitem[\protect\citeauthoryear{{Klypin}, {Yepes}, {Gottl{\"o}ber}, {Prada} \&
  {He{\ss}}}{{Klypin} et~al.}{2016}]{Klypin2016}
{Klypin} A.,  {Yepes} G.,  {Gottl{\"o}ber} S.,  {Prada} F.,    {He{\ss}} S.,
  2016, \mnras, 457, 4340

\bibitem[\protect\citeauthoryear{{Klypin}, {Trujillo-Gomez} \&
  {Primack}}{{Klypin} et~al.}{2011}]{Bolshoi}
{Klypin} A.~A.,  {Trujillo-Gomez} S.,    {Primack} J.,  2011, \apj, 740, 102

\bibitem[\protect\citeauthoryear{{Kofman}, {Bertschinger}, {Gelb}, {Nusser} \&
  {Dekel}}{{Kofman} et~al.}{1994}]{Kofman1994}
{Kofman} L.,  {Bertschinger} E.,  {Gelb} J.~M.,  {Nusser} A.,    {Dekel} A.,
  1994, \apj, 420, 44

\bibitem[\protect\citeauthoryear{{Kravtsov}, {Klypin} \& {Khokhlov}}{{Kravtsov}
  et~al.}{1997}]{ART}
{Kravtsov} A.~V.,  {Klypin} A.~A.,    {Khokhlov} A.~M.,  1997, \apjs, 111, 73

\bibitem[\protect\citeauthoryear{{Lam} \& {Sheth}}{{Lam} \&
  {Sheth}}{2008a}]{LamShethEllips2008}
{Lam} T.~Y.,  {Sheth} R.~K.,  2008a, \mnras, 389, 1249

\bibitem[\protect\citeauthoryear{{Lam} \& {Sheth}}{{Lam} \&
  {Sheth}}{2008b}]{LamSheth2008}
{Lam} T.~Y.,  {Sheth} R.~K.,  2008b, \mnras, 386, 407

\bibitem[\protect\citeauthoryear{{Lee}, {Primack}, {Behroozi},
  {Rodr{\'{\i}}guez-Puebla}, {Hellinger} \& {Dekel}}{{Lee}
  et~al.}{2017}]{Lee2017}
{Lee} C.~T.,  {Primack} J.~R.,  {Behroozi} P.,  {Rodr{\'{\i}}guez-Puebla} A.,
  {Hellinger} D.,    {Dekel} A.,  2017, \mnras, 466, 3834

\bibitem[\protect\citeauthoryear{{Manera} et~al.,}{{Manera}
  et~al.}{2013}]{Manera2013}
{Manera} M.  et~al., 2013, \mnras, 428, 1036

\bibitem[\protect\citeauthoryear{{Neyrinck}}{{Neyrinck}}{2016}]{Neyrinck2016}
{Neyrinck} M.~C.,  2016, \mnras, 455, L11

\bibitem[\protect\citeauthoryear{{Ohta}, {Kayo} \& {Taruya}}{{Ohta}
  et~al.}{2003}]{Ohta2003}
{Ohta} Y.,  {Kayo} I.,    {Taruya} A.,  2003, \apj, 589, 1

\bibitem[\protect\citeauthoryear{{Pandey}, {White}, {Springel} \&
  {Angulo}}{{Pandey} et~al.}{2013}]{Pandey2013}
{Pandey} B.,  {White} S.~D.~M.,  {Springel} V.,    {Angulo} R.~E.,  2013,
  \mnras, 435, 2968

\bibitem[\protect\citeauthoryear{{Peebles}}{{Peebles}}{1980}]{Peebles}
{Peebles} P.~J.~E.,  1980, {The large-scale structure of the universe}

\bibitem[\protect\citeauthoryear{{Platen}}{{Platen}}{2009}]{Platen2009}
{Platen} E.,  2009, PhD thesis, University of Groningen

\bibitem[\protect\citeauthoryear{{Sheth}}{{Sheth}}{1998}]{Sheth1998}
{Sheth} R.~K.,  1998, \mnras, 300, 1057

\bibitem[\protect\citeauthoryear{{Shin}, {Kim}, {Pichon}, {Jeong} \&
  {Park}}{{Shin} et~al.}{2017}]{Shin2017}
{Shin} J.,  {Kim} J.,  {Pichon} C.,  {Jeong} D.,    {Park} C.,  2017, ArXiv
  e-prints

\bibitem[\protect\citeauthoryear{{Soneira} \& {Peebles}}{{Soneira} \&
  {Peebles}}{1978}]{SoneiraPeebles1978}
{Soneira} R.~M.,  {Peebles} P.~J.~E.,  1978, \aj, 83, 845

\bibitem[\protect\citeauthoryear{{Springel}}{{Springel}}{2005}]{Gadget2}
{Springel} V.,  2005, \mnras, 364, 1105

\bibitem[\protect\citeauthoryear{{Takahashi}, {Oguri}, {Sato} \&
  {Hamana}}{{Takahashi} et~al.}{2011}]{Takahashi2011}
{Takahashi} R.,  {Oguri} M.,  {Sato} M.,    {Hamana} T.,  2011, \apj, 742, 15

\bibitem[\protect\citeauthoryear{{Taruya}, {Takada}, {Hamana}, {Kayo} \&
  {Futamase}}{{Taruya} et~al.}{2002}]{Taruya2002}
{Taruya} A.,  {Takada} M.,  {Hamana} T.,  {Kayo} I.,    {Futamase} T.,  2002,
  \apj, 571, 638

\bibitem[\protect\citeauthoryear{{Tassev}, {Zaldarriaga} \&
  {Eisenstein}}{{Tassev} et~al.}{2013}]{COLA}
{Tassev} S.,  {Zaldarriaga} M.,    {Eisenstein} D.~J.,  2013, Journal of
  Cosmology and Astroparticle Physics, 6, 036

\bibitem[\protect\citeauthoryear{{Wild} et~al.,}{{Wild}
  et~al.}{2005}]{Wild2005}
{Wild} V.  et~al., 2005, \mnras, 356, 247

\end{thebibliography}
\bibliographystyle{mn2e}
\appendix
 \makeatletter{}\section{Numerical effects: finite-volume variance,
  mass and force resolution}
\label{sec:cosmic}

\makeatletter{}\begin{figure*}
\centering
\includegraphics[width=0.495\textwidth]
{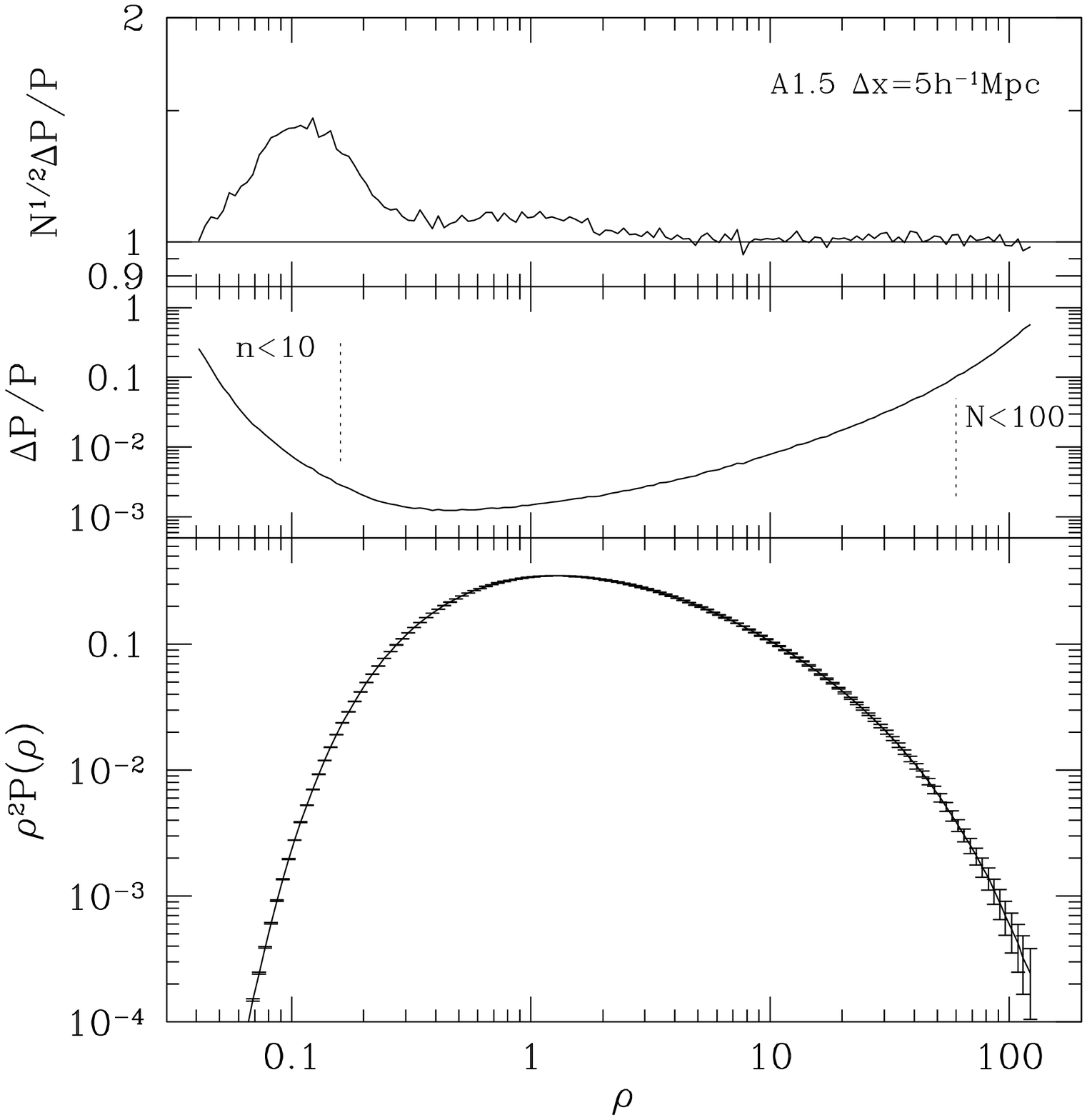}
\includegraphics[width=0.495\textwidth]
{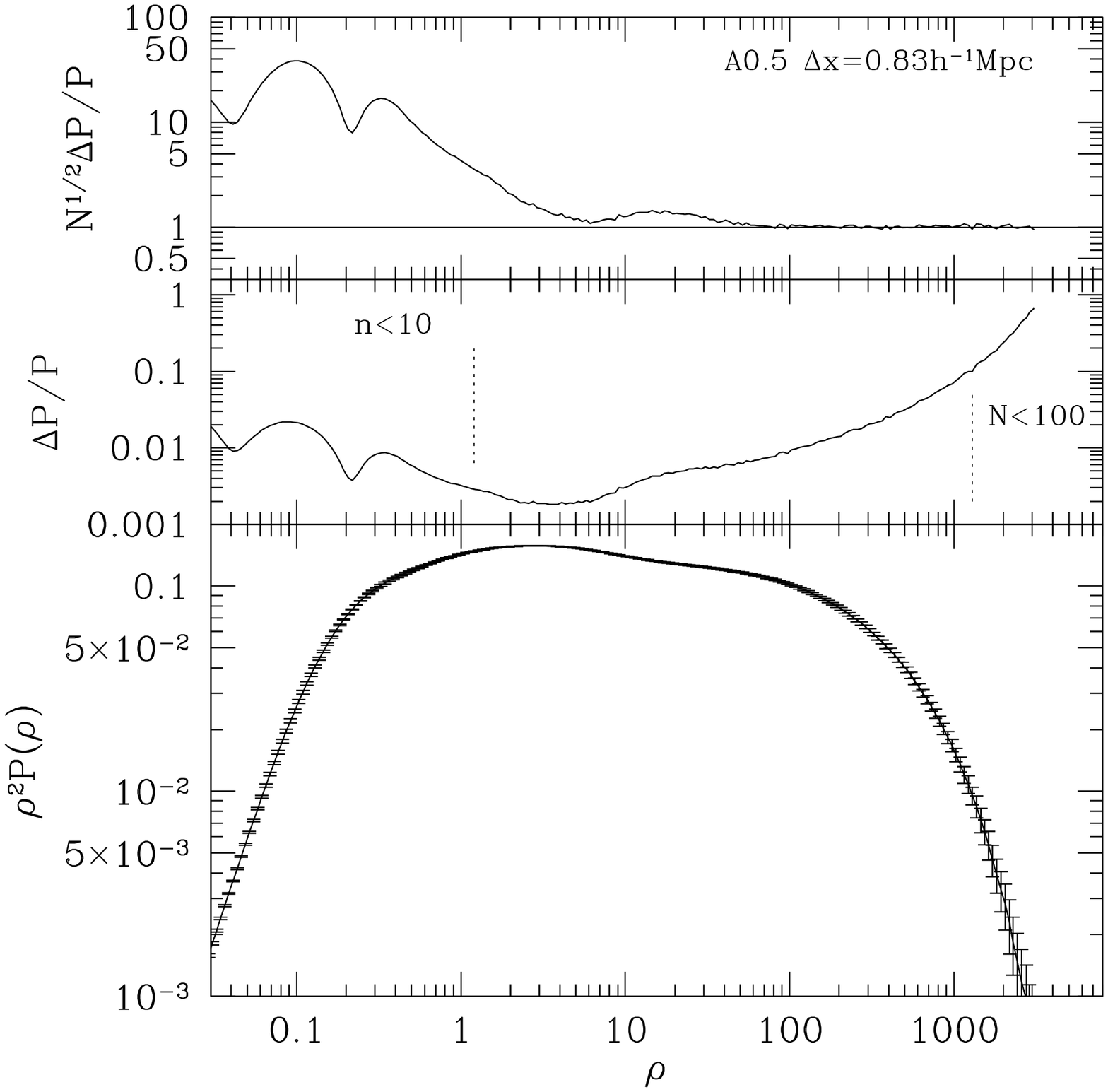}
\caption{Statistical errors of the density distribution function $P(\rho)$
   due to the finite-volume simulation variance.  Results are shown at $z=0$
  for the GLAM A1.5 (left) and A0.5 (right) simulations with different
  cell sizes. The middle panels present the $rms$ fluctuations
  $\Delta P/P$ of a single realization. The errors of the mean
  $P(\rho)$ are significantly lower because these simulations have
  very large number of realizations. The top panels show the
  $rms$ deviations scaled with $\sqrt{N}$, where $N$ is the number of
  cells in a bin of $P(\rho)$. For densities that are probed with a
  large number of particles the $rms$ fluctuations are defined by the
  number of cells per bin of $\rho$. The fluctuations are
  substantially non-Gaussian for bins with number of particles $n$ per bin less than 10.}
\label{fig:errors}
\end{figure*}

\makeatletter{}\begin{figure*}
\centering
\includegraphics[width=0.495\textwidth]
{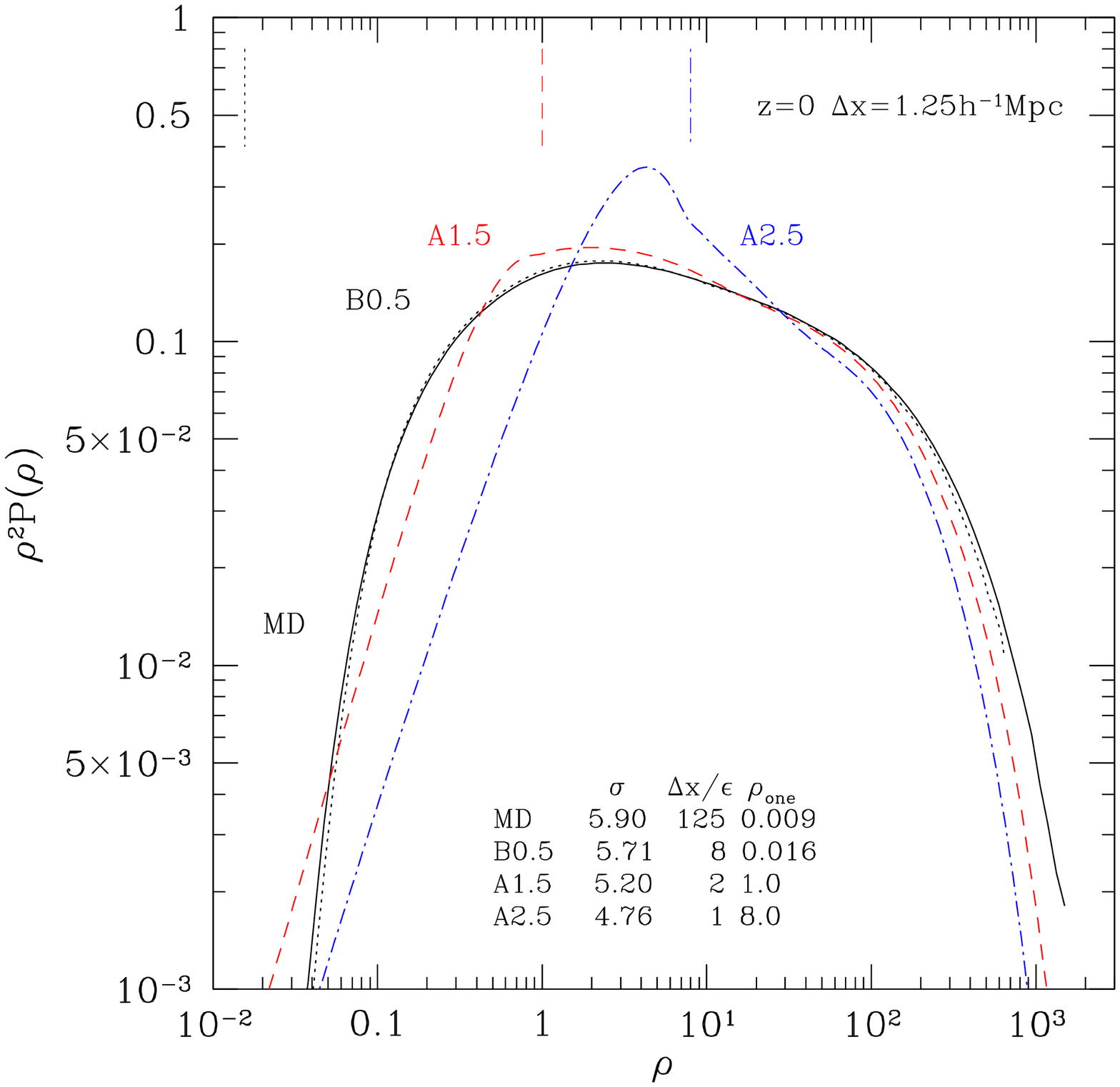}
\includegraphics[width=0.495\textwidth]
{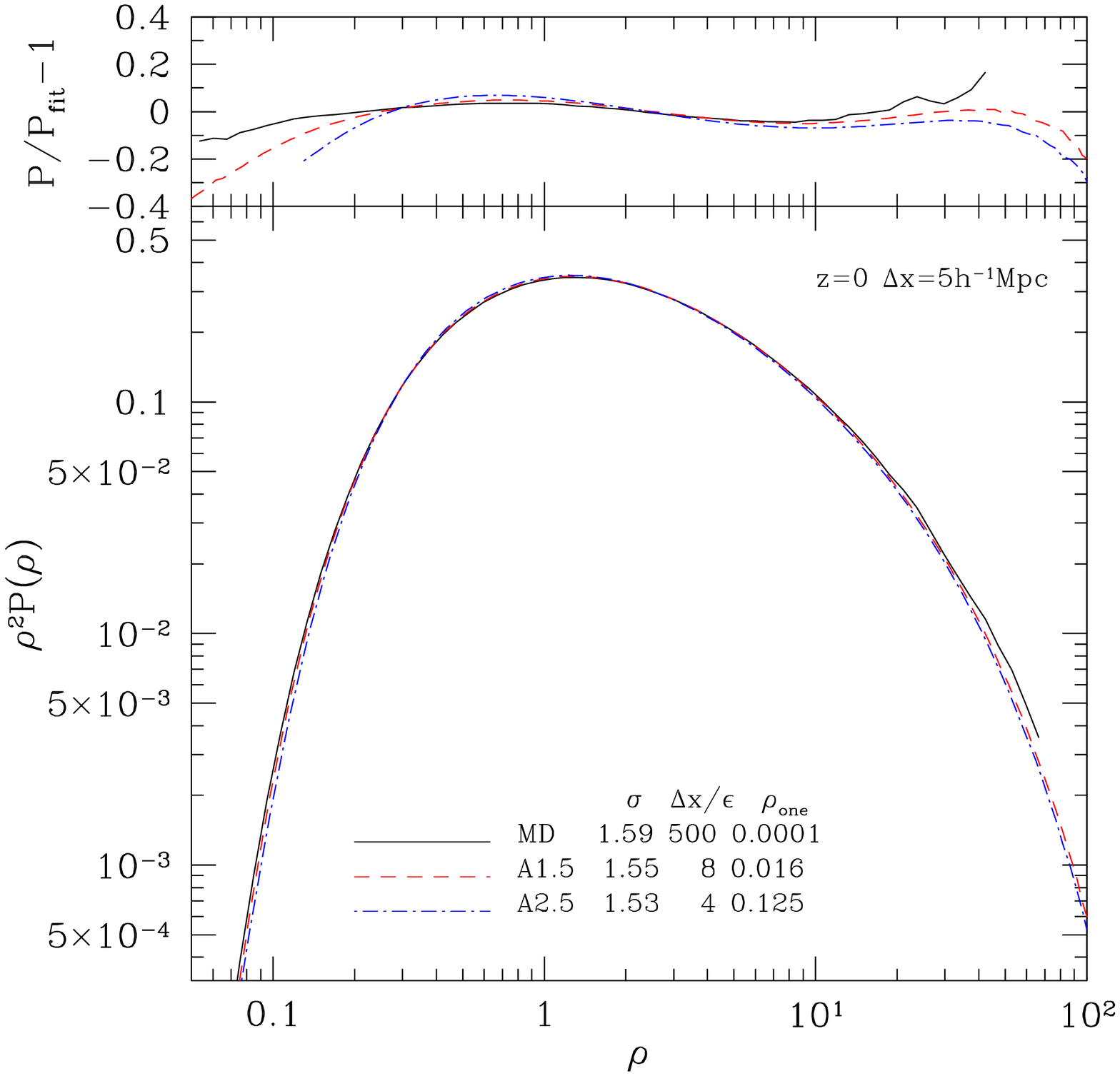}
\caption{Comparison of density distribution functions estimated in
  simulations with different box sizes and resolutions. {\it Left:}
  Example of numerical convergence for the cell size
  $\Delta x =1.25\Mpch$ at $z=0$. Insufficient force resolution
  reduces the amplitude of $P(\rho)$ at large densities.  As the force
  resolution increases, the results converge. In order to have errors
  less than few percent the ratio of the cell size $\Delta x$ to the
  force resolution $\epsilon$ should be larger than $\sim 8$. At small
  densities the noise due to the finite number-density of particles
  can produce large errors. The vertical lines at the top of the panel
  show the density $\rho_{\rm one}$ that a single particle produces
  when placed at the center of the cell. The bump in the A2.5 curve at
  $\rho \lesssim \rho_{\rm one}$ is due to large discreteness effects:
  for this cell size the A2.5 simulations did not have enough
  particles.  {\it Right:} The same as for the left panel, but now for
  better force resolution $\Delta x/\epsilon$ and for larger
  number-densities of particles.  To facilitate detailed comparisons
  the top panel show deviations of $P(\rho)$ from the same analytical
  fit. Results indicate that the box size does not affect the density
  distribution. However, the force resolution has a tendency to reduce
  the amplitude of $P(\rho)$ at very large density $\rho$.  }
\label{fig:boxsize}
\end{figure*}

Numerical effects may play an important role for the estimates of the
density distribution function.  We start with the analysis of the effects of
the variance due to the fine-volume of the simulations. We use the
GLAM simulations A0.5 with small cell
$\Delta x=0.83\Mpch$ and A1.5 with larger cell $\Delta x=0.5\Mpch$ to
estimate the level of fluctuations in different realizations. Bottom
panels in Figure~\ref{fig:errors} show the average values of
$\rho^2P(\rho)$ and the statistical fluctuations of {\it a single
  realization}. As expected, the fluctuations due to the fine-volume simulation
variance are larger for very large densities and become very small for
$\rho\approx 0.5-10$.

The middle panels present relative fluctuations $\Delta P/\langle P\rangle$,
where $\langle P\rangle$ is the average PDF over an ensamble of realizations and
$\Delta P$ is the $rms$ deviation. The vertical dotted lines at large
$\rho$ show the density bin with 100 cells in a single realization. At this
density the level of statistical fluctuations $\Delta P/P$ is about
0.1, which is consistant with the expected shot noise. We clarify this
situation by plotting in the top panels the relative fluctuations
scaled with $N^{1/2}$, where $N$ is the number of cells of given
density in a bin.  Indeed, the fluctuations  are  defined
by the number of cells in a bin for large densities.

The situation is different at small densities $\rho\lesssim 1$ where
the fluctuations become substantially stronger than the Gaussian
$\Delta P/P = N^{-1/2}$. This is likely related with the increasing
noise in the density field due to too few particles per cell $n$. The
vertical dotted lines on low densities show the bin at which $n=10$.

In spite of being strongly non-Poissonian at small densities, the errors
are still very small. For example, at $\rho=0.1$ for the A0.5 simulations
the errors are just $\sim 2$\% for a single realization (right
panels).  Note that the fluctuations plotted in
Figure~\ref{fig:errors} provide average deviations of a single
realization from the ensamble average. Errors of the average $P(\rho)$
are significantly smaller. For example, for the A1.5 simulations (left
panels) the error of the mean at $\rho=60$, $N=100$ is just
$\sim 0.1$\%.

In summary, our results are mostly dominated by systematics, not
by the variance. When dealing with individual simulations such as
MDPL1 or BolshoiP we use only bins with more than $N>100$ per bin. For
large sets of simulations A0.5, 1.5, 2.5 we accept bins with more than
10 cells.

In order to evaluate other possible numerical effects, we select two
filtering scales $\Delta x=1.25\Mpch$ and $\Delta x=5\Mpch$ and
analyze $P(\rho)$ at $z=0$ obtained from different simulations. The two
filtering scales probe different dynamical regimes. For
$\Delta x=1.25\Mpch$ the $rms$ density fluctuation
 is large $\sigma \approx 5$. So, we are testing
very nonlinear regime with densities up to 1000.  The larger filtering
scale $\Delta x=5\Mpch$ probes more modest fluctuations with
$\sigma \approx 1.5$ and densities $\rho =0.1 - 100$.

By comparing these results we test the effects of the finite box size
(ranging from $500\Mpch$ to $2500\Mpch$), the effects of force
resolution $\Delta x/\epsilon$ (ranging from 1 to 500) and 
the discreteness effects associated with the finite number of dark
matter particles. The later can be characterized by the density
$\rho_{\rm one}$ produced by a single particle placed at the node of a
cell:
\begin{equation}
\rho_{\rm one} = \left(\frac{L}{\Delta x N_p}\right)^3,
\end{equation}
where $N_p^3$ is the number of particles and  $L$ is the size of the simulation box.
Vertical lines in the left panel of Figure~\ref{fig:boxsize}
show $\rho_{\rm one}$ for different simulations.  Values of different 
parameters are also given in Figure~\ref{fig:boxsize}.

We use the left panel in Figure~\ref{fig:boxsize} to demonstrate two
numerical effects. At large densities $\rho \gtrsim 50$ the
discreteness effects are small even for the A2.5 simulations and the
results are dominated by the force resolution. Here we find a trend
that is expected for simulations with low force resolution: the PDF
increases with increasing force resolution. However, there is
little difference between simulations when the resolution becomes
sufficiently high: the PDF for B0.5 simulations with 8 resolution
elements across one cell( $\Delta x/\epsilon=8$) is nearly
the same as for the much high resolution simulation MDPL1 with
$\Delta x/\epsilon=125$. This is a signature that the results have
converged.

The discreteness of density assignment becomes an issue at small
densities as is clearly seen in Figure~\ref{fig:boxsize} for
$\rho\lesssim 10$. The A2.5 simulations provide a good example on how the
particle noise affects the PDF. There is a large bump in
$\rho^2P(\rho)$ at densities slightly below $\rho_{\rm one}$. At much
smaller densities $\rho\approx 0.1\rho_{\rm one}$ the PDF falls much
below the real one. The effects of the particle noise extend to
densities above $\rho_{\rm one}$ but quickly die out beyond
$\rho\approx 10\rho_{\rm one}$.  
\label{sec:all}

The right panel in Figure~\ref{fig:boxsize} shows better convergence
of $P(\rho)$ because we select simulations that have
better force resolution and smaller level of the particle noise. In
order to see the differences more clearly we make a fit to the data,
and on the top panel plot we show only the deviations from the fit. The A2.5
data still fall below the more accurate MDPL1 results both at small densities
(effects of the particle noise) and large densities (effects of the force
resolution). The A1.5 data show much smaller errors.

Using the results presented in Figure~\ref{fig:boxsize} and similar
results of the comparison between different simulations (e.g., comparison of
BolshoiP and MDPL1) we find limits on numerical parameters that should be
satisfied to produce a PDF $P(\rho)$ with errors less than a few percent: 

\begin{itemize}
\item the filtering scale $\Delta x$ must be resolved with not less than 8
force resolution elements: $\Delta x> 8\epsilon$
\item  the number of particles per
filtering cell should be not less than 10-20: $\rho>(10-20)\rho_{\rm one}$.
\end{itemize}

\newpage

\section{Parameters of the double-exponential model for the Dark Matter Density Distribution Function}

Tables~\ref{tab:fit0} and~\ref{tab:fit1} give parameters for the
double-exponential model given in eq.(\ref{eq:dexp}) for
different simulations, filtering scales and redshifts.

\begin{table*}
  \caption{Parameters for the double-exponential model, eq.(\ref{eq:dexp}), for redshift $z=0$. Different columns give: 
    (1) Name of the simulation, (2) Cell size in $\Mpch$, (3) $rms$ density fluctuation, (4-7) parameters for eq.(\ref{eq:dexp}),
    (8) $rms$ of the relative error of the fit, (9) number of cells in 1D, (10) relative force resolution -- the number of force resolution elements per each density assignment cell.}
\begin{center}
  \tabcolsep 7.2pt
\begin{tabular*}{0.88\textwidth}{@{}lccccccccr@{}}
\hline\hline
Simulation & Cell Size & $\sigma$ & $A$ & $\alpha+2$& $\rho_0$ & $\rho_1$ & relative error  & $N_{\rm cell}$ & $\Delta x/\epsilon$ \\
(1)  & (2)     & (3)     & (4)         & (5)      & (6)      & (7)     & (8)    & (9) & (10)  \\
\hline
MDPL1   &   0.167 & 24.020  & 1.145E-01   &  0.0438  &  0.1001  & 2080.000 &  0.066 & 6000 &  17 \\
MDPL1   &   0.250 & 19.070  & 1.122E-01   &  0.0639  &  0.0925  & 1225.927 &  0.058 & 4000 &  25 \\
MDPL1   &   0.313 & 16.630  & 1.148E-01   &  0.0677  &  0.1104  &  894.088 &  0.056 & 3200 &  31 \\
MDPL1   &   0.333 & 15.960  & 1.175E-01   &  0.0670  &  0.1351  &  810.545 &  0.057 & 3000 &  33 \\
MDPL1   &   0.500 & 12.150  & 1.306E-01   &  0.0640  &  0.1359  &  451.211 &  0.056 & 2000 &  50 \\
MDPL1   &   0.625 & 10.340  & 1.385E-01   &  0.0636  &  0.1238  &  318.285 &  0.052 & 1600 &  62 \\
A0.5 &   0.833 &  7.753  & 1.651E-01   &  0.0457  &  0.1518  &  180.093 &  0.052 & 600 & 4\\
MDPL1   &   1.000 &  7.140  & 1.730E-01   &  0.0441  &  0.1523  &  143.084 &  0.045 & 1000 & 100 \\
MDPL1   &   1.250 &  5.902  & 1.983E-01   &  0.0274  &  0.1754  &   95.484 &  0.040 &  800 & 125 \\
B0.5 &   1.250 &  5.712  & 2.059E-01   &  0.0050  &  0.1868  &  103.010 &  0.033 & 400 & 8 \\
MDPL1   &   1.429 &  5.243  & 2.178E-01   &  0.0123  &  0.1946  &   75.355 &  0.036 &  700 & 143 \\
A0.5 &   1.667 &  4.422  & 2.519E-01   & -0.0232  &  0.2281  &   58.198 &  0.026 & 300 & 8\\
MDPL1   &   2.000 &  3.834  & 2.843E-01   & -0.0185  &  0.2488  &   36.196 &  0.030 &  500 & 200 \\
MDPL1   &   2.500 &  3.102  & 3.498E-01   & -0.0399  &  0.2952  &   21.230 &  0.027 &  400 & 250 \\
B0.5 &   2.500 &  3.048  & 3.587E-01   &  -0.0455 &   0.3021 &   20.465 &  0.026 & 200 & 16 \\
A0.5 &   3.333 &  2.299  & 5.024E-01   & -0.0545  &  0.3756  &    9.010 &  0.022 & 150 & 16\\  
MDPL1   &   3.333 &  2.343  & 4.852E-01   & -0.0520  &  0.3672  &    9.604 &  0.027 &  300 & 333 \\
A1.5 &   5.000 &  1.555  & 9.813E-01   & -0.0590  &  0.5324  &    2.829 &  0.020 &  300 & 8\\
MDPL1   &   5.000 &  1.586  & 8.902E-01   & -0.0500  &  0.4950  &    3.100 &  0.020 &  200 & 500 \\
A1.5 &  10.000 &  0.836  & 8.976E+00   & -0.0423  &  1.0600  &    0.361 &  0.017 & 150 & 16\\
A2.5 &  10.000 &  0.832  & 9.127E+00   & -0.0477  &  1.0971  &    0.368 &  0.013 & 250 & 8 \\
A2.5 &  20.000 &  0.452  & 2.198E+03   & -0.0340  &  2.5080  &    0.052 &  0.069 & 125 & 16 \\
\hline\hline
\end{tabular*}
\end{center}
\label{tab:fit0}
\end{table*}

\begin{table*}
  \caption{Parameters for the double-exponential model, eq.(\ref{eq:dexp}), for redshifts $z\approx 0.5-1$. Different columns give: 
    (1) Name of the simulation, (2) Cell size in $\Mpch$, (3) $rms$ density fluctuation, (4-7) parameters for eq.(\ref{eq:dexp}),
    (8) $rms$ of the relative error of the fit, (9) number of cells in 1D, (10) relative force resolution -- the number of force resolution elements per each density assignment cell, (11) redshift.}
\begin{center}
  \tabcolsep 7.2pt
\begin{tabular*}{0.96\textwidth}{@{}lcccccccccr@{}}
\hline\hline
Simulation & Cell Size & $\sigma$ & $A$ & $\alpha+2$& $\rho_0$ & $\rho_1$ & relative error  & $N_{\rm cell}$ & $\Delta x/\epsilon$ & redshift \\
(1)  & (2)     & (3)     & (4)         & (5)      & (6)      & (7)     & (8)    & (9) & (10)  & (11) \\
\hline
MDPL1 &    0.167 &  16.050 &   1.448E-01 &   0.0146 &  0.0714  &  946.937  &  0.062 & 6000 & 16.7 & 0.492 \\
MDPL1 &    0.250 &  12.780 &   1.392E-01 &   0.0346 &  0.0685  &  555.436  &  0.048 & 4000 & 25.0 & 0.492 \\
MDPL1 &    0.500 &   8.177 &   1.683E-01 &   0.0218 &  0.1562  &  213.103  &  0.045 & 2000 & 50.0 & 0.492 \\
MDPL1 &    1.000 &   4.793 &   2.476E-01 &  -0.0461 &  0.2313  &   72.823  &  0.031 & 1000 & 100. & 0.492 \\
MDPL1 &    1.250 &   3.961 &   2.909E-01 &  -0.0680 &  0.2668  &   46.150  &  0.028 &  800 & 125. & 0.492 \\
B0.5 &  1.250 &  3.837  &   2.993E-01 & -0.0806  &  0.2764  &  45.377   &  0.026 &  400 & 8.   & 0.492 \\
A0.5 &  1.667 &  2.923  &  3.882E-01  &  -0.0816 &  0.3300  &   19.242  &  0.038 &  300 & 8    & 0.5097 \\
B0.5 &  2.500 &  2.068  &   6.078E-01 & -0.0673  &  0.4228  &   6.282   &  0.042 &  200 & 16   & 0.492 \\
MDPL1 &    2.500 &   2.112 &   5.801E-01 &  -0.0690 &  0.4130  &    6.892  &  0.044 &  400 & 250. & 0.492 \\
A0.5 &  3.333 &  1.575  &  1.015E+00  &  -0.0616 &  0.5398  &    2.635  &  0.038 &  150 & 16   & 0.5097 \\
MDPL1 &    5.000 &   1.138 &   2.413E+00 &  -0.0538 &  0.7508  &    0.974  &  0.033 &  200 & 500. & 0.492 \\
A2.5 &  10.00 &  0.625  &  7.163E+01  & -0.0526  & 1.6335   &    0.141  &  0.017 & 250  & 8    & 0.497 \\
A2.5 &  20.00 &  0.347  &  1.409E+05  & -0.0261  & 3.5731   &    0.024  &  0.140 & 125  & 16   & 0.497 \\
B0.5 &  1.25 &  2.671  & 4.516E-01 & -0.0597 & 0.3612 & 11.71  & 0.092  &  400 & 8 & 0.99 \\
A0.5 &  1.667&  2.066 &  6.462E-01 & -0.0541 & 0.4374 & 5.372  & 0.090  & 300 &8  & 0.99 \\
B0.5 &  2.50 &  1.475  & 1.248E+00 & -0.0501 & 0.5876 & 1.966  & 0.069  &  200 & 16 & 0.99 \\
A0.5 &  3.333&  1.157 &  2.484E+00 & -0.0460 & 0.7563 & 0.941  & 0.056  & 150 &16 & 0.99 \\
A1.5 &  5.00 &  0.833 &  1.013E+01 & -0.0413 & 1.1148 & 0.343  & 0.043  & 300 & 8  & 1.0 \\
A1.5 & 10.00 &  0.487 &  1.134E+03 & -0.0436 & 2.3475 & 0.061  & 0.032  & 150 & 16 & 1.0 \\
A2.5 & 10.00 &  0.485 &  1.297E+03 & -0.0403 & 2.4207 & 0.0597 & 0.0564 & 250 & 8  & 1.0 \\
A2.5 & 20.00 &  0.274 &  3.522E+09 & -0.0319 & 6.1154 & 0.0079 & 0.0782 & 125 & 16 & 1.0 \\
\hline\hline
\end{tabular*}
\end{center}
\label{tab:fit1}
\end{table*}

\end{document}